\documentclass{article}
\usepackage{tcolorbox}
\usepackage{geometry}
\usepackage{amsmath}
\usepackage{graphicx}
\usepackage{wasysym}
\usepackage{natbib}
\usepackage{amssymb}
\usepackage[hidelinks]{hyperref}
\usepackage{url}
\usepackage{array}
\usepackage{subcaption}
\usepackage{gensymb}
\usepackage{algpseudocode}
\usepackage{soul}
\usepackage{multirow}
\usepackage{listings}
\usepackage{xcolor}
\usepackage{dsfont}
\usepackage{float}
\usepackage{amsfonts}
\usepackage{graphicx}
\usepackage{caption}
\usepackage{subcaption}
\usepackage{tikz}
\usepackage{wrapfig}
\usepackage{listings}
\usepackage{xcolor}

\definecolor{codegreen}{rgb}{0,0.6,0}
\definecolor{codegray}{rgb}{0.5,0.5,0.5}
\definecolor{codepurple}{rgb}{0.58,0,0.82}
\definecolor{backcolour}{rgb}{1,1,1}
\definecolor{navy}{rgb}{0.0, 0.0, 0.5}
\definecolor{darkgreen}{rgb}{0.0, 0.5, 0.0}

\lstdefinestyle{mystyle}{
    backgroundcolor=\color{backcolour},   
    commentstyle=\color{codegreen},
    keywordstyle=\color{blue},
    numberstyle=\tiny\color{codegray},
    stringstyle=\color{codepurple},
    basicstyle=\ttfamily\footnotesize,
    breakatwhitespace=\text{false},         
    breaklines=\text{true},                 
    captionpos=b,                    
    keepspaces=\text{true},                 
    numbers=left,                    
    numbersep=5pt,                  
    showspaces=\text{false},                
    showstringspaces=\text{false},
    showtabs=\text{false},                  
    tabsize=4
}
\title{Boids of a Feather Flock Together — Evolving Prey Behaviours Under Different Predator Attack Strategies} 

\author{Augusta van Haren \\ {\normalsize\itshape Radboud University} \and Hanna Hoogen \\ {\normalsize\itshape Radboud University} \and Luca Pattavina \\ {\normalsize\itshape Radboud University}}

\date{\today \\[6pt] \normalsize\itshape{All authors contributed equally.}}

\begin{document}
\maketitle

\begin{abstract}
\noindent Flocking and schooling are thought to have evolved partly as defences against predation, but how prey should balance social and escape tendencies may depend on the predator's hunting strategy. We extend the predator--prey boids model of \citet{Ojo_2023}, itself based on Reynolds' boids, by combining six prey movement tendencies (alignment, cohesion, separation, dodge, repel and wiggle) into a single weighted acceleration update, and by reformulating wiggle as a sinusoidal manoeuvre. We then use an evolutionary strategy to optimise the six behaviour coefficients for collective prey survival against four predator hunting strategies: attack-centroid, attack-nearest, attack-random and attack-peripheral. Across five independent trials per strategy, coefficients converged within trials and mean fitness remained stable or increased, although trials often settled in different local optima. Prey survival was highest under attack-centroid and lowest under attack-nearest, in line with our hypotheses. Against attack-centroid, prey evolved individualistic predator avoidance with high escape coefficients, whereas against the other three strategies they largely kept their flock formation. Across all strategies, evolution favoured a low repel coefficient and relatively high dodge and wiggle coefficients. Our results suggest that optimal anti-predator behaviour depends on the interplay between escape tendencies and the predator's hunting strategy.
\end{abstract}

\section{Introduction}

Among many lifeforms, complex collective behaviour can emerge from simple social interactions, as seen in flocks of birds, schools of fish, and insect colonies. In a group, individuals influence one another and adapt their behaviour relying only on local neighbourhood information. However, such behaviour does not take place in isolation from the environment; together they form a complex, dynamical system. For example, individuals of different species or groups can interact and form predator and prey relationships. Both predators and prey have to evolve dynamic, adaptive behaviours to survive. Collective behaviours, such as schooling and flocking, may have evolved to decrease predation losses by confusing the predator or collectively organising information within a group of agents with limited information \citep{Kunz2006-ze}. These collective behaviours can be governed by simple local rules. A classic example is the boids model by \cite{reynolds1987flocks}. This model simulates realistic flocking, which can be used to learn about flock behaviour in different scenarios through artificial boid simulations. One such scenario is the presence of a predator, where the trade-off between behavioural tendencies becomes crucial for survival \citep{palmer2021reactive, lee2006dynamics}.

In this work, we evolve optimal prey escape behaviour based on local rules, under different predator attack strategies. To evolve the optimal behaviours, we again look at nature -- algorithms inspired by evolution are useful for solving optimisation problems without relying on gradients \citep{beyer2002evolution, slowik2020evolutionary}. Here, we use an evolutionary strategy (ES) to optimise continuous behaviour coefficients.

\subsection{Research Question}
The aim of this work is to implement an extended version of \citeauthor{Ojo_2023}'s predator-prey model \citeyearpar{Ojo_2023}, which is based on \citeauthor{reynolds1987flocks}'s boids model \citeyearpar{reynolds1987flocks}. We enhance this previous work by combining six prey behaviour tendencies (alignment, cohesion, separation, dodge, repel, wiggle) into one behaviour vector that governs the boids' acceleration. Further, we apply an ES to find the optimal coefficients for the behaviour tendencies for four different predator attack strategies. Moreover, we make small improvements to the behaviour implementations. With this setup, we then aim to answer the following research question:
\begin{center}
    \emph{What movement behaviour (in terms of relative contributions of alignment, cohesion, separation, dodge, repel and wiggle tendencies) do prey evolve to, in order to maximise the collective survival of their population under different predator hunting strategies?}
\end{center}
We hypothesise that specific combinations of prey behaviour coefficients are optimal under different predator strategies, leading to distinct flocking and escape dynamics. Further, we expect the prey to achieve the highest survival rate under the ``attack-centroid'' predator strategy, as the centroid does not necessarily correspond to a specific boid, making the strategy inefficient \citep{Demsar2014-km}. Moreover, we hypothesise that prey achieve the lowest survival rate under the ``attack-nearest'' strategy, as this strategy is simple but effective \citep{vonmoll2016evolutionary}.

\subsection{Related Work}
\citeauthor{reynolds1987flocks}'s \citeyearpar{reynolds1987flocks} boids model provides the fundamental structure for simulating flocking behaviour and agent-movement, as governed by three principles: alignment, cohesion, and separation. We further build on the extension of the Reynolds' boids model of \citet{Ojo_2023}, who implemented two classes of boids: predator and prey. \citet{Ojo_2023} further provide four predator attack strategies (attack-centroid, attack-nearest, attack-random, attack-peripheral) and three escape behaviours (dodge, repel, wiggle), which govern the prey's behaviour together with the classic boid principles. While relying heavily on the simulation implementation by \citet{Ojo_2023}, we make several key improvements. Most crucially, \citet{Ojo_2023} analyse the prey's flocking behaviour together with one of the prey escape tendencies in isolation. 
Instead, we combine all three behaviours together with the flocking principles into one acceleration update vector -- allowing the prey to make use of all escape tendencies simultaneously, and learn the most effective behaviour for a given predator's hunting strategy. Furthermore, in analogy with \citeauthor{reynolds1987flocks}'s \citeyearpar{reynolds1987flocks} boids model, we treat this prey's acceleration update vector as a weighted sum of all of its components -- whereas \citeauthor{Ojo_2023}'s \citeyearpar{Ojo_2023} implementation did not assign distinct weights for every movement tendency. Moreover, we further improve their wiggle behaviour, which we describe in more detail in Section \ref{methods:boidsmodel}, and performed sensitivity analyses on key simulation parameters. 

Other studies, similar to \citet{Ojo_2023} investigated predator and prey strategies, like \citep{Demsar2014-km} who explored the effect of different predator strategies (attack-centroid, attack-nearest, attack-peripheral) on social and individualistic prey and found flocking to be the optimal anti-predatory behaviour -- they further characterised the escape patterns exhibited by the prey, similar to \citet{lee2006dynamics}. 

The models described so far, relied on pre-specified behaviours and strategies with fixed parameters. We instead aim to learn prey behaviours using an ES \citep{beyer2002evolution, slowik2020evolutionary}. Previous studies similarly attempted to learn predator or prey behaviours; for example, \citet{alaliyat2022optimization} used a genetic algorithm (GA) to evolve the boids coefficients to achieve realistic flocking behaviour. Similarly, \citet{Hahn2019-kd} used reinforcement learning in a predator-prey simulation to learn the prey's strategy, which resulted in flocking behaviour similar to the boids model to confuse the predator. Furthermore, \citet{Kunz2006-ze} used an evolutionary algorithm to evolve prey behaviour while varying predator parameters. \citet{oboshi2002evolving} used a GA to evolve prey evasion behaviour for one attack strategy (attack-nearest) in a simulation setup similar to ours. However, all of these studies operated on a more basic level than ours, by focusing on learning general collective behaviours like flocking and swarming -- instead of specific escape strategies to avoid a predator. 

Few studies tried to evolve more sophisticated behaviours in a predator-prey scenario; \citet{vonmoll2016evolutionary} focused on evolving predator strategies in a scenario with two predators by using a GA. Similarly to our work, they evolved coefficients for a fixed set of atomic behaviours (pursue, converge, diverge, drive, flank) -- however, targeting predators rather than prey. Moreover, \citet{chen2006genetic} use a GA to evolve coefficients in a linear combination of behaviours, consisting of the classic boid principles, as well as obstacle avoidance, following feed, and avoiding a predator. However, they only implement a direction-based escape behaviour (corresponding to ``dodge'') and do not present any proper results. 

To summarise, our work extends the existing literature by using an \emph{evolutionary strategy (ES)} to perform continuous optimisation on the coefficients for \emph{a combination of prey behaviours} to avoid predation under \emph{different predator attack strategies}.

\section{Methods}
All analyses were performed using Python 3.11.9 on Windows. The code, result files, and sample simulation videos are available under \url{https://github.com/ivychad/NC-Project-Code-Boids}. We provide an in-depth explanation for the choice of all fixed parameters in Appendix \ref{app:fixed_parameter_settings}.

\subsection{Boids Model}
\label{methods:boidsmodel}
To study our research question, we adapted the flocking model of \cite{Ojo_GitHub}. This implementation is an extension of \citeauthor{reynolds1987flocks}'s boids model \citeyearpar{reynolds1987flocks} -- which dictates that the movement dynamics of each boid are based on three rules: alignment, cohesion, and separation. \cite{Ojo_GitHub} extend this model by defining two types of boids, predator and prey $B\in\{B^{pred},B^{prey}\}$, that each have distinct rules which govern their movement behaviour. Furthermore, they add realism to various aspects of the original model. We have refined their approach by combining the six movement tendencies into one acceleration update vector, by enhancing realism of the wiggle tendency, and by evolving the optimal coefficients for the relative contribution of each of those tendencies using an ES.

The implementation equips every boid $B$ with a position $\mathbf{p}(B)\in\mathbb{R}^2$ and velocity $\mathbf{v}(B)\in\mathbb{R}^2$. In each iteration of the simulation, the imposed acceleration $\mathbf{a}(B)\in\mathbb{R}^2$ on the boid determines its dynamics of motion: 
\begin{align*}
    \mathbf{v}(B) &\ \leftarrow \ \mathbf{v}(B) + \mathbf{a}(B)\cdot dt\\
    \mathbf{p}(B) &\ \leftarrow \ \mathbf{p}(B) + \mathbf{v}(B) \cdot dt + \frac{1}{2} \ \mathbf{a}(B)\cdot dt^2 
\end{align*}
In the traditional boids model, updates in the boid's acceleration and velocity are effectively unbounded. As this is physically unrealistic, \citeauthor{Ojo_GitHub}'s \citeyearpar{Ojo_GitHub} implementation limits the direction as well as magnitude of acceleration. This implies that if the angle between the boid's current velocity $\textbf{v}(B)$ and the imposed acceleration $\textbf{a}(B)$ exceeds the maximum rotation angle ($|\angle\big({\hat{\mathbf{a}}}(B),{\hat{\mathbf{v}}}(B)\big)|>\theta_{max}^{pred,prey}$), the updated acceleration vector is limited to $\mathbf{\hat{a}}(B)= \textbf{R}(\pm\theta_{max}^{pred,prey})\ \mathbf{\hat{v}}(B)$ in the given direction ($\textbf{R}$ is the rotation matrix), and rescaled to $||\mathbf{a}(B)|| =a^{pred,prey}$ \citep{Ojo_2023}. Together, this limits the turn speed of the boid. Moreover, the magnitude of velocity of the boid is bounded to $||\mathbf{v}(B)||=v^{pred,prey}$. All bounds are specific to the prey or predator dynamics, and manually adjusted to model realistic behaviour, see Appendix \ref{app:fixed_parameter_settings}.

Lastly, all boids are assigned a (predator- and prey-specific) perception radius $r_P^{pred,prey}$, separation radius $r_S^{pred,prey}$, and a perception angle $fov^{pred,prey}$. For each movement tendency, these parameters determine which other boids $B_i$ are `seen', and considered a `neighbour' of $B$. In the traditional boids model, this only depends on Euclidean distance $\text{\text{dist}}(B,B_i)=||\mathbf{p}(B),\mathbf{p}(B_i)||_{2}<{r_{P,S}^{pred,prey}}^2$), as if both predators and preys had a field of view of 360°. However, to enhance realism, \cite{Ojo_GitHub} also requires neighbours to be within $B$'s visual field ($\text{inFov}(B,B_i)=\text{true}$). This condition entails that $B_i$ is within $B$'s field of view (delimited by $fov^{pred,prey}$). In addition, $B_i$ should not be \text{occluded} by another boid ($ \text{occluded}(B,B_i)=\text{false}$). In this definition, neighbour $B_i$ occludes $B_j$ if they are separated by less than $2\degree$ in $B$'s visual field, and $B_i$ is closer than $B_j$. This results in excluding $B_j$ to be a neighbour of $B$ \citep{Ojo_2023}.

\subsubsection{Predator}
For a predator boid $B:=B^{pred}$ with corresponding predator parameters, its acceleration $\mathbf{a}(B)$ is governed by either one of the following hunting strategies \citep{Ojo_GitHub, Ojo_2023}. These strategies are in line with the attack strategies proposed by \cite{Demsar2014-km}:
$    \mathbf{\hat{a}}(B) \in  \{
    \mathbf{\hat{a}_{attc}}(B),   \mathbf{\hat{a}_{attn}}(B),
    \mathbf{\hat{a}_{attr}}(B),
    \mathbf{\hat{a}_{attp}}(B)\}.
$ Its acceleration and velocity are limited as described above. The hunting strategies each characterise a specific movement behaviour:

\begin{itemize}
    \item \textbf{Attack-centroid}: $B$ targets the average position of all its neighbouring prey boids $B^{prey}_i$\\
    \hspace*{.3cm} $\mathbf{\hat{a}_{attc}}(B) \ = \ \frac{\sum_{i=1}^{N_B} (\mathbf{p}(B^{prey}_i) - \mathbf{p}(B))}{N_B}$\\
    \hspace*{.8em} where each of the $N_B$ neighbouring prey boids $B^{prey}_i$ satisfy \\
    \hspace*{.8em} $\text{dist}(B,B^{prey}_i) < {r_P^{pred}}^2 \ \ \wedge \ \  \text{inFov}(B,B^{prey}_i)=\text{true} \ \ \wedge \ \  \text{occluded}(B,B^{prey}_i)=\text{false}$

    \item \textbf{Attack-nearest}: $B$ targets the position of the nearest neighbouring prey boid $B^{prey}_t$\\
    \hspace*{.3cm} $\mathbf{\hat{a}_{attn}}(B) \ = \ \mathbf{p}(B^{prey}_t) - \mathbf{p}(B)$\\
    \hspace*{.8em} where $B^{prey}_t = \arg \min_{B^{prey}_i} \text{dist}(B,B_i^{prey\displaystyle })$ is closest to $B$ out of the set\\ 
    \hspace*{.8em} of $N_B$ neighbouring prey boids of which each $B_i^{prey}$ satisfies\\
    \hspace*{.8em} $\text{dist}(B,B^{prey}_i) < {r_P^{pred}}^2 \ \ \wedge \ \  \text{inFov}(B,B^{prey}_i)=\text{true} \ \ \wedge \ \  \text{occluded}(B,B^{prey}_i)=\text{false}$\
    
    \item \textbf{Attack-random}: $B$ targets the position of a random neighbouring prey boid $B^{prey}_t$\\
    \hspace*{.3cm} $\mathbf{\hat{a}_{attr}}(B) \ = \ \mathbf{p}(B^{prey}_t) - \mathbf{p}(B)$\\
    \hspace*{.8em} where $B^{prey}_t \sim \{B_1^{prey},\dots,B_{N_B}^{prey}\}$ is chosen randomly from the set of $N_B$ neighbouring prey boids of\\
    \hspace*{.8em} which each $B^{prey}_i$ satisfies \\
    \hspace*{.8em} $\text{dist}(B,B^{prey}_i) < {r_P^{pred}}^2 \ \ \wedge \ \  \text{inFov}(B,B^{prey}_i)=\text{true} \ \ \wedge \ \  \text{occluded}(B,B^{prey}_i)=\text{false}$

    \item \textbf{Attack-peripheral}: $B$ targets the position of the most peripheral prey boid $B^{prey}_t$\\
    \hspace*{.3cm} $\mathbf{\hat{a}_{attp}}(B) \ = \ \mathbf{p}(B^{prey}_{t}) - \mathbf{p}(B)$\\
    \hspace*{.8em} where $B^{prey}_{t} = \arg \max_{B^{prey}_i} \big(
    \frac{\sum_{j=1}^{N_B}\mathbf{p}(B^{prey}_j)}{N_B} -
    \mathbf{p}(B_i^{prey})
    \big)$ is most isolated from the centroid\\
    \hspace*{.8em} of the $N_B$ neighbouring prey boids of which each $B_i^{prey}$ satisfies\\
    \hspace*{.8em} $\text{dist}(B,B^{prey}_i) < {r_P^{pred}}^2 \ \ \wedge \ \  \text{inFov}(B,B^{prey}_i)=\text{true} \ \ \wedge \ \  \text{occluded}(B,B^{prey}_i)=\text{false}$\
\end{itemize}

\subsubsection{Prey}
For a prey boid $B:=B^{prey}$ with corresponding prey parameters, its acceleration $\mathbf{a}(B)$ constitutes a superposition of multiple movement tendencies. As an extension upon \cite{Ojo_GitHub}'s implementation, we combine all movement tendencies as a weighted sum into an acceleration update vector. Each tendency's sub-acceleration $\hat{\mathbf{a}}_{\square}$ is weighed by a corresponding coefficient $c_{\square}$, hereby contributing to the (normalised) direction of acceleration:
\begin{align*}
\hat{\mathbf{a}}(B) \ = \
{norm}\big(
    c_{ali} \cdot \hat{\mathbf{a}}_{ali}(B) +
    c_{coh} \cdot \hat{\mathbf{a}}_{coh}(B) + 
    c_{sep} \cdot \hat{\mathbf{a}}_{sep}(B)\ + \\
    \quad \quad \ \ \
    c_{dod} \cdot \hat{\mathbf{a}}_{dod}(B) +
    c_{rep} \cdot \hat{\mathbf{a}}_{rep}(B) +
    c_{wig} \cdot \hat{\mathbf{a}}_{wig}(B)
\big)
\end{align*}
Its acceleration and velocity are limited as described above. The first three movement tendencies give rise to the boids flocking behaviour \citep{Ojo_2023, Ojo_GitHub, reynolds1987flocks}:

\begin{itemize}
    \item \textbf{Alignment}: $B$ matches the average direction and speed of all its neighbouring boids $B^{prey}_i$ (consistent direction flocking)\\
        \hspace*{.3cm} $\mathbf{\hat{a}_{ali}}(B) \ = \ norm\left(\frac{\sum_{i=1}^{N_B} \mathbf{v}(B^{prey}_i)}{N_B} - \mathbf{v}(B)\right)$\\
        \hspace*{.8em} where each of the $N_B$ neighbouring prey boids $B^{prey}_i$ satisfy \\
        \hspace*{.8em} $\text{dist}(B,B^{prey}_i) < {r_P^{prey}}^2 \ \ \wedge \ \  \text{inFov}(B,B^{prey}_i)=\text{true} \ \ \wedge \ \  \text{occluded}(B,B^{prey}_i)=\text{false}$
        
    \item \textbf{Cohesion}: $B$ matches the average position of all its neighbouring prey boids $B^{prey}_i$ (localised flocking)\\
    \hspace*{.3cm} $\mathbf{\hat{a}_{coh}}(B) \ = \ norm\left(\frac{\sum_{i=1}^{N_B} (\mathbf{p}(B^{prey}_i) - \mathbf{p}(B))}{N_B}\right)$\\
    \hspace*{.8em} where each of the $N_B$ neighbouring prey boids $B^{prey}_i$ satisfy \\
    \hspace*{.8em} $\text{dist}(B,B^{prey}_i) < {r_P^{prey}}^2 \ \ \wedge \ \  \text{inFov}(B,B^{prey}_i)=\text{true} \ \ \wedge \ \  \text{occluded}(B,B^{prey}_i)=\text{false}$

    \item \textbf{Separation}: $B$ repels the positions of all its neighbouring prey boids $B^{prey}_i$ (preventing collisions)\\
    \hspace*{.3cm} $\mathbf{\hat{a}_{sep}}(B)\  = \ norm\left(\sum_{i=1}^{N_B} (\mathbf{p}(B) - \mathbf{p}(B^{prey}_i))\right)$\\
    \hspace*{.8em} where each of the $N_B$ neighbouring prey boids $B^{prey}_i$ satisfy\\
    \hspace*{.8em} $\text{dist}(B,B^{prey}_i) < {r_S^{prey}}^2 \ \ \wedge \ \  \text{inFov}(B,B^{prey}_i)=\text{true} \ \ \wedge \ \  \text{occluded}(B,B^{prey}_i)=\text{false}$
\end{itemize}
The remaining tendencies characterise escape manoeuvres to flee from the predator. We implemented three distinct escape behaviours \citep{Ojo_2023,Ojo_GitHub}, as first introduced in the empirically-based Homing Pigeons Escape (HoPE) \citep{Papadopoulou2022-fn}:
\begin{itemize}
    \item \textbf{Dodge}: $B$ dodges approaching predator boids $B^{pred}_i$, by turning perpendicularly in the opposite direction (direction-based escaping)\\
    \hspace*{.3cm}
    ${\mathbf{\hat{a}_{dod}}}(B)\ =\ norm\left(\sum_{i=1}^{M_B} 
    \left\{
    \begin{array}{ll}
    \textbf{R}(+90\degree)\ \hat{\mathbf{v}}(B) & \text{if } \angle\big(\hat{\mathbf{v}}(B), \hat{\mathbf{v}}(B^{{pred}}_i)\big) \leq 0\degree\\
    \textbf{R}(-90\degree)\ \hat{\mathbf{v}}(B) & \text{if } \angle\big(\hat{\mathbf{v}}(B), \hat{\mathbf{v}}(B^{{pred}}_i)\big) > 0\degree
    \end{array}
    \right.\right)$\\
    \hspace*{.8em} where each of the $M_B$ neighbouring predator boids $B^{pred}_i$ satisfy\\
    \hspace*{.8em} $\text{dist}(B,B^{pred}_i) < {r_P^{prey}}^2 \ \ \wedge \ \  \text{inFov}(B,B^{pred}_i)=\text{true} \ \ \wedge \ \  \text{occluded}(B,B^{pred}_i)=\text{false}$

    \item \textbf{Repel}: $B$ repels the positions of all approaching predator boids $B^{pred}_i$ (position-based escaping)\\
    \hspace*{.3cm} $\mathbf{\hat{a}_{rep}}(B)\  = \ norm\left(\sum_{i=1}^{M_B} (\mathbf{p}(B) - \mathbf{p}(B^{pred}_i))\right)$\\
    \hspace*{.8em} where each of the $M_B$ neighbouring predator boids $B^{pred}_i$ satisfy\\
    \hspace*{.8em} $\text{dist}(B,B^{pred}_i) < {r_P^{prey}}^2 \ \ \wedge \ \  \text{inFov}(B,B^{pred}_i)=\text{true} \ \ \wedge \ \  \text{occluded}(B,B^{pred}_i)=\text{false}$

    \item \textbf{Wiggle}: $B$ wiggles by some wiggle angle $\theta_{wig}$ and frequency $f_{wig}$ if a predator is near (shake-off escaping)\\
    \hspace*{.3cm} $\mathbf{\hat{a}_{wig}}(B)\  =
    \begin{cases}
        \textbf{R}(\theta_{wig},f_{wig})\ \mathbf{\hat{v}}(B) & \text{if $M_B>0$}\\
        \mathbf{\hat{v}}(B) & \text{if $M_B= 0$}\\
    \end{cases}$\\
    \hspace*{.8em} where each of the $M_B$ neighbouring predator boids $B^{pred}_i$ satisfy\\
    \hspace*{.8em} $\text{dist}(B,B^{pred}_i) < {r_P^{prey}}^2 \ \ \wedge \ \  \text{inFov}(B,B^{pred}_i)=\text{true} \ \ \wedge \ \  \text{occluded}(B,B^{pred}_i)=\text{false}$
\end{itemize}
We enhanced \citep{Ojo_GitHub}'s wiggle behaviour implementation by redefining the wiggle in terms of a sine wave with specific angle and frequency, to resemble more naturalistic behaviour.

\subsubsection{Simulation}
To analyse the emerging movement behaviour of the boids, a field of $S_x\times S_y$ is initialised with $M$ predators and $N$ prey. Governed by their predator- and prey-specific parameters and coefficients, movement behaviour will naturally emerge. When a predator `catches' a prey ($\text{dist}(B_i^{pred},B_j^{prey})\leq (r_S^{pred})^2 $), this prey is removed from the field.

\subsection{Evolutionary Strategy (ES)}
To study the research question at hand, we constructed an Evolutionary Strategy (ES) \citep{beyer2002evolution, slowik2020evolutionary} around the aforementioned simulation. This ES consists of $\mathcal{N}_g$ generations, of which each generation runs $\mathcal{N}_s$ simulations -- defining the population size. Each simulation then denotes one individual of the population, and each simulation is run for a simulation time of $\mathcal{T}$ steps -- having a \emph{fixed} number of predators ($M$) and prey ($N$), a \emph{fixed} set of predator and prey parameters, and a \emph{fixed} predator hunting strategy $\mathbf{\hat{a}}(B)\in\big\{\mathbf{\hat{a}_{attc}}(B),\mathbf{\hat{a}_{attn}}(B),\mathbf{\hat{a}_{attr}}(B),\mathbf{\hat{a}_{attp}}(B)\big\}$.

Crucially, the prey coefficients are \emph{variable} across simulations, and thereby characterise the movement dynamics of a unique simulation. Therefore, we define the `gene' of a simulation $s$ as their prey coefficients, $gene(s):=\langle {c_{ali}}_s,{c_{coh}}_s,{c_{sep}}_s,{c_{dod}}_s,{c_{rep}}_s,{c_{wig}}_s\rangle$.

\subsubsection{Fitness}
The fitness of a simulation $s$, denoted $f(s)$, is defined from the prey's perspective, and reflects the number of surviving prey $N_{\mathcal{T}}\leq N$ at the end of simulation (after $\mathcal{T}$ time-steps):
\begin{equation*}
    f(s)=N_{\mathcal{T}}
\end{equation*}

\subsubsection{Generation Updating}
Given a generation $g_t$ of $\mathcal{N}_s$ simulations (individuals), the two fittest simulations are automatically carried over to the next generation $g_{t+1}$ -- ensuring that the most successful combination of coefficients are not lost (elitism, where elite size $\mu_e=2$). To create the remaining $\mathcal{N}_s-2$ individuals, repeatedly two `parents' are selected by means of fitness-proportional parent selection, i.e., sampled where simulation $s$ has a probability to be selected of:
\begin{equation*}
    p(s)=\frac{f(s)}{\sum_{i=1}^{\mathcal{N}_s}f(s_i)}
\end{equation*}
Given two sampled parents $s_i,s_j$ of generation $g_t$, two children $s'_{i'},s'_{j'}$ are created to populate generation $g_{t+1}$. We employ crossover to explore the parameter space, where a random crossover point $c$ defines how the parents' genes are recombined:
\begin{align*}
gene(s'_{i'})&=gene(s_i)_{[:c]}+gene(s_j)_{[c:]}\\
gene(s'_{j'})&=gene(s_j)_{[:c]}+gene(s_i)_{[c:]}
\end{align*}
Furthermore, mutation adds a small random value $\epsilon_{\square}\in[-0.1,0.1]$ to each prey coefficient $c_{\square}\in gene(s'_{i'}),gene(s'_{j'})$, by mutation rate $\mu$:
\begin{equation*}
    c_{\square} \leftarrow
        \begin{cases}
        c_{\square} + \epsilon_{\square} & \text{with probability } \mu \\
        c_{\square} & \text{with probability } 1 - \mu
        \end{cases}
\end{equation*}
The resulting prey coefficients are clipped to $0\leq c_{\square}\leq1$ to keep them within the allowed range. This reproduction process is repeated for $\frac{\mathcal{N}_s-2}{2}$ times per generation update -- hereby ensuring that the next generation again counts $\mathcal{N}_s$ individuals.

\subsection{Experimental Design}
\label{experimentaldesign}
To answer our research question, we regarded the predator's hunting strategy $\mathbf{\hat{a}}(B)$ as the independent variable, and the distribution of prey coefficients in $gene(s):=\langle {c_{ali}}_s,{c_{coh}}_s,{c_{sep}}_s,{c_{dod}}_s,{c_{rep}}_s,{c_{wig}}_s\rangle$ over generations as the dependent variables of interest -- characterising the evolution of the prey's movement behaviour.

For each hunting strategy $\mathbf{\hat{a}}(B)\in\big\{\mathbf{\hat{a}_{attc}}(B),\mathbf{\hat{a}_{attn}}(B),\mathbf{\hat{a}_{attr}}(B),\mathbf{\hat{a}_{attp}}(B)\big\}$, we ran the ES for $\mathcal{N}_g=30$ generations, keeping all other parameters except the prey coefficients fixed according to Table \ref{tab:fixed_parameter_settings}. For the first generation $g_1$, we initialised all $\mathcal{N}_s=30$ simulations individually with random prey coefficients, i.e., $gene(s)\sim\mathcal{U}(0,1)^6$. This generation was then evolved over $\mathcal{N}_g=30$ generations, while recording all simulations' prey coefficients $gene(s):=\langle c_{ali}, c_{coh}, c_{sep}, c_{dod}, c_{rep}, c_{wig}\rangle$, as well as their fitnesses $f(s)$. These coefficients were then aggregated to assess their distributions over generations -- hereby assessing the prey's evolutionary response to distinct hunting strategies. Over generations, we expected the prey coefficients to converge to their optimal values, and as a result, the mean simulation fitness (collective survival rate) to rise. To account for stochasticity (while bearing our computational limitations in mind), for each predator hunting strategy we repeated this ES trial five times. 

The optimal parameter settings for the main experiments (see Table \ref{tab:fixed_parameter_settings}) were largely found by trial-and-error, as motivated in Appendix \ref{app:fixed_parameter_settings}. We scrutinised $r_P^{prey}$, $fov^{prey}$, $\mathcal{N}_s$ and $\mu$ into more detail, as these parameters showed to have most impact in the stability of the ES. In tuning these parameters, our goal was to identify configurations that led to both an increase in collective fitness over generations and stable convergence of the prey coefficients. As we note that simulation results may heavily depend on these fixed parameter settings, the impact of varying these parameters was systematically assessed through the sensitivity analyses (see Appendix \ref{app:sensitivity_analyses}).

\section{Results}
\label{results}
\noindent By conducting the experiments described in subsection \ref{experimentaldesign}, we recorded the dynamics of the prey coefficients -- $gene(s):=\langle c_{ali}, c_{coh}, c_{sep}, c_{dod}, c_{rep}, c_{wig}\rangle$ -- along with the fitnesses $f(s)$ of each simulation run (i.e., ES individual) over 30 generations -- yielding 31 values (initialisation included). For each predator hunting strategy, this procedure was repeated in five independent ES trials to reduce the impact of stochasticity. The resulting dynamics were aggregated to assess the prey's evolutionary response to different hunting strategies.

\begin{figure}[h!]
\centering
\begin{tabular}{cc}
  \subcaptionbox{Attack-centroid hunting strategy, $\mathbf{\hat{a}}(B)=\mathbf{\hat{a}_{attc}}(B)$\label{fig:fitness_centroid}}{
    \includegraphics[width=0.31\textwidth]{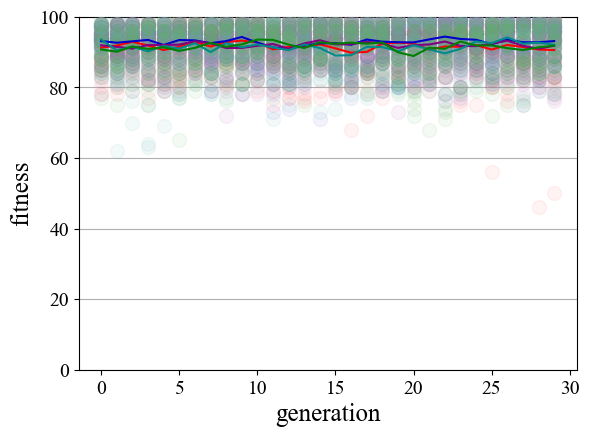}
  } &
  \subcaptionbox{Attack-nearest hunting strategy, $\mathbf{\hat{a}}(B)=\mathbf{\hat{a}_{attn}}(B)$\label{fig:fitness_nearest}}{
    \includegraphics[width=0.31\textwidth]{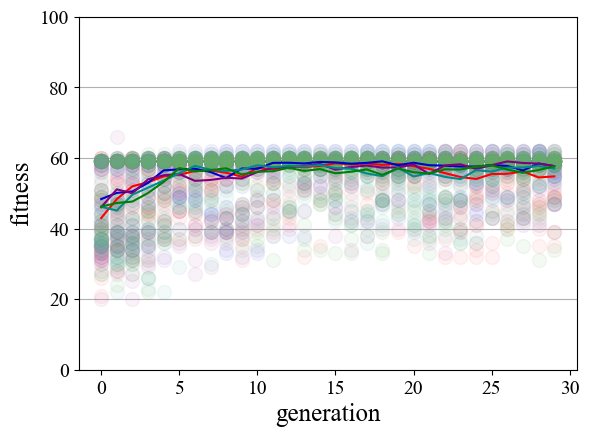}
  } \\
  \subcaptionbox{Attack-random hunting strategy, $\mathbf{\hat{a}}(B)=\mathbf{\hat{a}_{attr}}(B)$\label{fig:fitness_random}}{
    \includegraphics[width=0.31\textwidth]{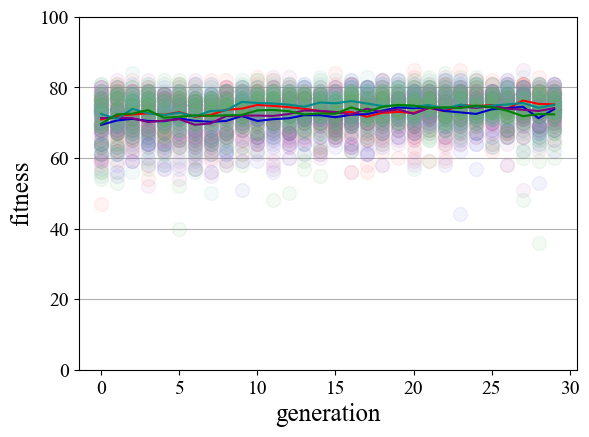}
  } &
  \subcaptionbox{attack-peripheral hunting strategy, $\mathbf{\hat{a}}(B)=\mathbf{\hat{a}_{attp}}(B)$\label{fig:fitness_peripheral}}{
    \includegraphics[width=0.31\textwidth]{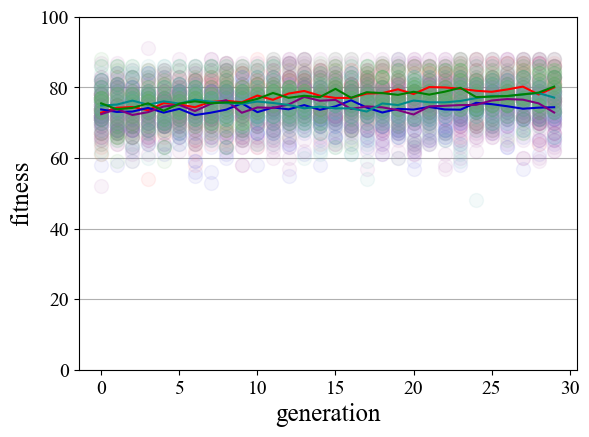}
  }
\end{tabular}
\includegraphics[width=0.7\textwidth]{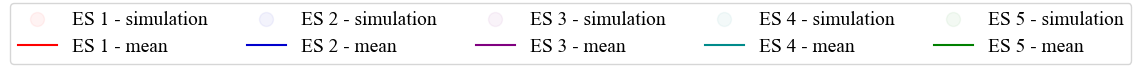}
\caption{Evolution of the distribution of fitnesses $f(s)$ over generations, for five different ES trials (corresponding to five different colours), per predator hunting strategy. Each dot corresponds to an individual simulation, where overlapping simulations result in darker-shaded dots. The line represents the population's mean fitness. Note that for every ES trial, the mean fitness remains stable or rises over generations.}
\label{fig:fitness_all}
\end{figure}

Generally speaking, the mean population fitness remained relatively stable (see Figure \ref{fig:fitness_centroid}), or showed a very slight (see Figure \ref{fig:fitness_random} and \ref{fig:fitness_peripheral}) to more pronounced (see Figure \ref{fig:fitness_nearest}) increase. As no ES trial showed a downwards nor heavily fluctuating trend of mean fitness, this confirms the desired functioning of the ES. In other words, the prey managed to learn to improve their set of prey coefficients $gene(s)$ over generations, leading to movement behaviour that enhanced their collective survival rate -- within the fitness limitations set by the predator's hunting strategy $\mathbf{\hat{a}}(B)\in\big\{\mathbf{\hat{a}_{attc}}(B),\mathbf{\hat{a}_{attn}}(B),\mathbf{\hat{a}_{attr}}(B),\mathbf{\hat{a}_{attp}}(B)\big\}$, which we discuss in detail below. Furthermore, we note that per hunting strategy, the mean fitness trends of the five ES trials agree -- indicating that the observed collective fitness improvements are dependent on, and characteristic to, the employed hunting strategy. 

In the remainder of this section, we present the results relating to each individual hunting strategy in more detail. Apart from reflecting upon the strategy-specific fitness trends, we will also assess the trends per prey coefficient -- which are visualised in Figure \ref{fig:coefficients_centroid}, \ref{fig:coefficients_nearest}, \ref{fig:coefficients_random} and \ref{fig:coefficients_peripheral}. For these, in general, we noted that \emph{within} each ES trial, the distributions of prey coefficients converge over generations. This convergence can be attributed to the ES progressing towards a local optimum. As each ES trial shows convergent behaviour of the coefficients, we therefore conclude that the coefficients' landscape allows sufficient opportunities for fitness improvement. However, this coefficients' landscape seems to contain multiple local optima -- as demonstrated by some ES trials that tend to bifurcate as they unfold -- resulting in the ES converging to two distinct values. This points at the coefficients' convergence being inter-dependent.

Regarding the prey coefficients' agreement \emph{between} ES trials of the same hunting strategy, on the other hand, convergence patterns do not always agree. Here we observe that, depending on the hunting strategy, some prey coefficients show convergence of ES trials towards the \emph{same} local optima -- indicating that the coefficients' landscape is sufficiently steep to globally converge, and suggesting that the coefficient is rather crucial for the collective survival rate of the prey. For other coefficients, however, different ES converge towards different local optima -- exhibiting more variability and a sparser distribution, and possibly indicating a less pronounced or more context-dependent influence on prey survival. The collected results for each hunting strategy will be assessed into more detail in the following subsections, while a comparative and comprehensive analysis, along with our interpretation for our findings, is provided in the discussion section \ref{Discussion}. 

\subsection{Attack-Centroid Hunting Strategy ($\mathbf{\hat{a}}(B)=\mathbf{\hat{a}_{attc}}(B)$)}

\begin{figure}[h!]
    \centering
    \includegraphics[width=\linewidth]{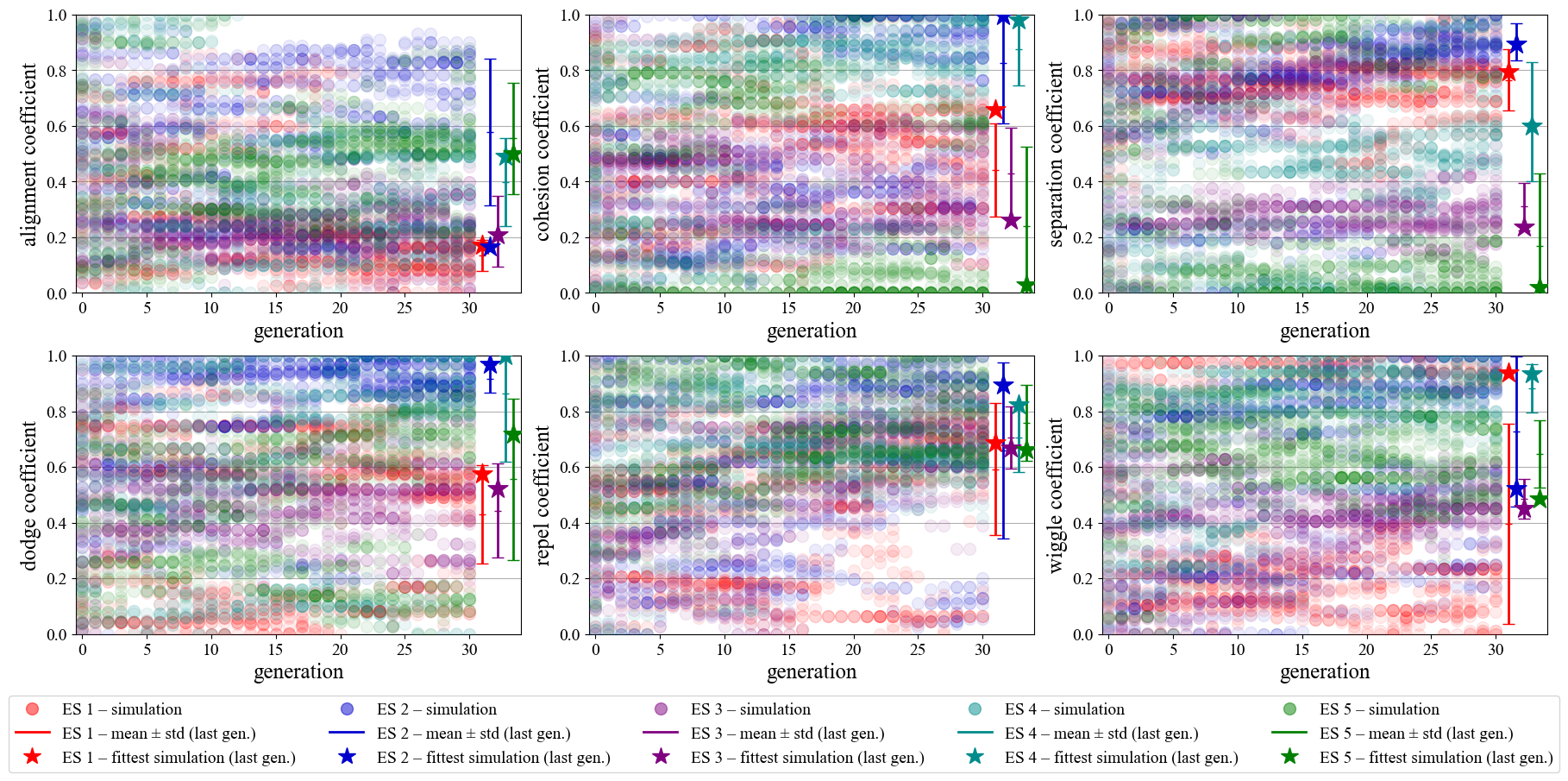}
    \caption{From left to right, top to bottom: evolution of prey coefficients $gene(s):=\langle c_{ali}, c_{coh}, c_{sep}, c_{dod}, c_{rep}, c_{wig}\rangle$ under the attack-centroid hunting strategy, $\mathbf{\hat{a}}(B)=\mathbf{\hat{a}_{attc}}(B)$, for five different ES trials -- corresponding to five different colours. Each dot (\CIRCLE) corresponds to an individual simulation, where overlapping simulations result in darker-shaded dots. The vertical bars indicate the population's error margin (mean $\pm$ std) at the last generation of each ES, whereas the stars ($\bigstar$) indicate the coefficient's value of the fittest individual of the last generation.}
    \label{fig:coefficients_centroid}
\end{figure}

\noindent With predators attacking the centroid of the prey flock ($\mathbf{\hat{a}}(B)=\mathbf{\hat{a}_{attc}}(B)$), we can observe from Figure \ref{fig:fitness_centroid} how the initial generations' fitnesses already start at near-maximal fitnesses -- which is sustained as each ES unfolds. In other words, the predator's attack-centroid strategy results in little prey to be caught, with little dependence on the prey's movement dynamics. 

Figure \ref{fig:coefficients_centroid} shows that coefficients tend to converge \emph{within} their ES run -- reflected by their decreased standard deviation at the last generation. Furthermore, the coefficient  distributions are generally centred around the coefficient's value of the fittest individual -- indicating that fitness improvements were the driving force behind the convergence of the ES. It should be noted, however, that convergence does not always agree \emph{between} ES trials. Whereas we see that the escape tendencies, i.e., dodge ($c_{dod}$), repel ($c_{rep}$) and wiggle ($c_{wig}$), all tend to converge towards higher values in $[0.4,1.0]$ -- an effect which is most pronounced for the repel coefficient, as indicated by the highest agreement of error bars between ES trials. 

For the flocking tendencies, i.e., alignment ($c_{ali}$), cohesion ($c_{coh}$) and separation ($c_{sep}$), however, the ES shows little agreement of convergence \emph{between} ES trials -- indicated by the lack of overlap of error margins at the last generation. Taken together, these findings suggest that individual fitness improvements are mostly realised by avoiding the predator, as opposed to changing the flock's global dynamics; this is also evident from simulation runs using the evolved coefficients for the five ES runs. Visual inspection of the fittest simulation of each ES trial revealed that for the attack-centroid strategy the prey mainly stays around the edges of the toroidal environment, while the predator moves around the centre of the environment which coincides with the centroid of the flock (given the toroidal environment). Prey consistently react to the predator by changing their direction while generally keeping the flock formation (albeit considerably less than for all other attack strategies).

\subsection{Attack-Nearest Hunting Strategy ($\mathbf{\hat{a}}(B)=\mathbf{\hat{a}_{attn}}(B)$)}

\begin{figure}[h!]
    \centering
    \includegraphics[width=\linewidth]{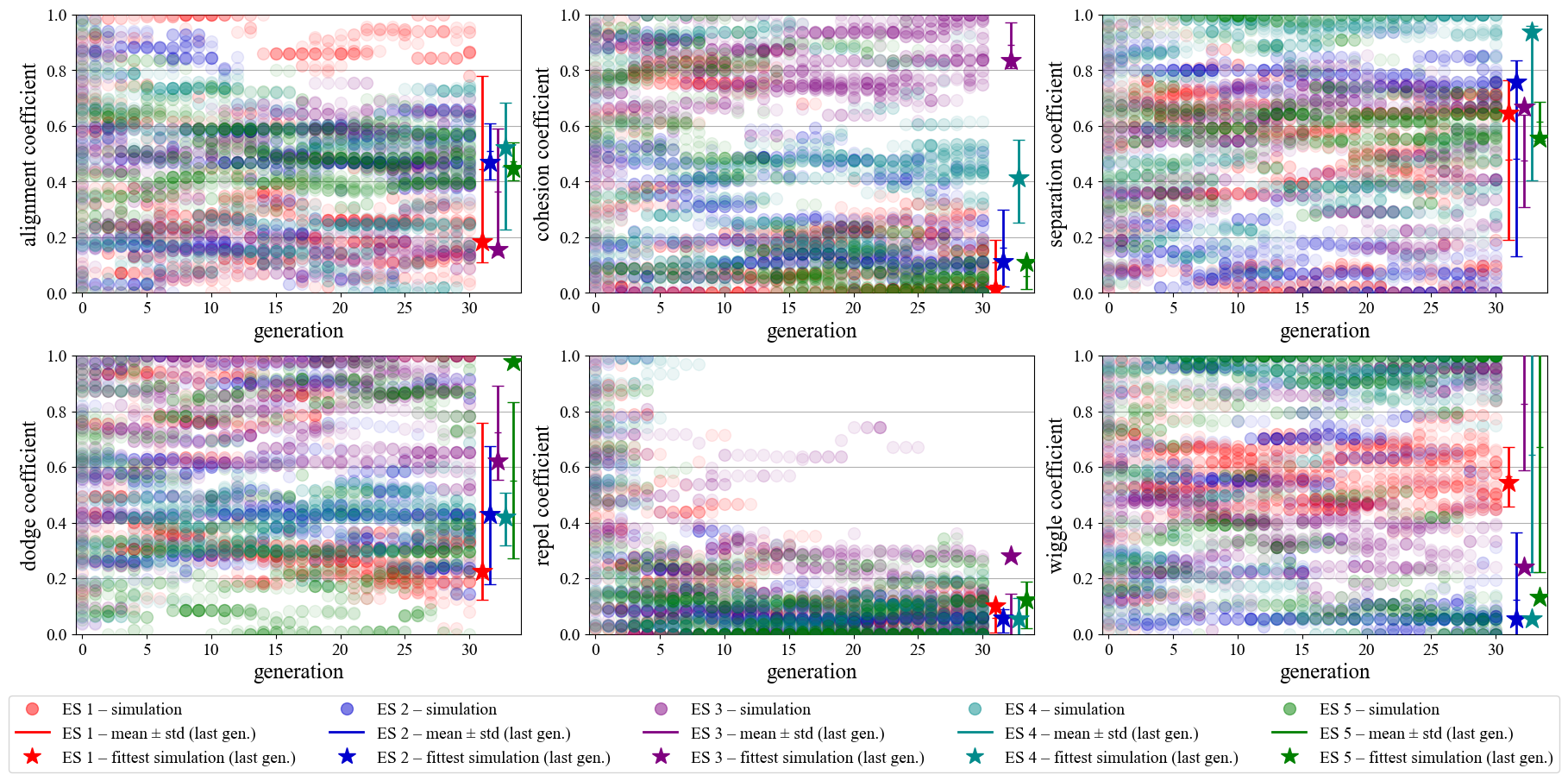}
    \caption{From left to right, top to bottom: evolution of prey coefficients $gene(s):=\langle c_{ali}, c_{coh}, c_{sep}, c_{dod}, c_{rep}, c_{wig}\rangle$ under the attack-nearest hunting strategy, $\mathbf{\hat{a}}(B)=\mathbf{\hat{a}_{attn}}(B)$, for five different ES trials -- corresponding to five different colours. Each dot (\CIRCLE) corresponds to an individual simulation, where overlapping simulations result in darker-shaded dots. The vertical bars indicate the population's error margin (mean $\pm$ std) at the last generation of each ES, whereas the stars ($\bigstar$) indicate the coefficient's value of the fittest individual of the last generation.}
    \label{fig:coefficients_nearest}
\end{figure}

\noindent When the predator attacks the nearest prey in the flock ($\mathbf{\hat{a}}(B)=\mathbf{\hat{a}_{attn}}(B)$), we can observe from Figure \ref{fig:fitness_nearest} how the escape for the prey is generally more difficult than in the previous attack-centroid scenario. Indeed, the initial fitness in the first generations is relatively low -- showing a survival rate lower than 50\% -- and, despite steadily increasing in the next ES generations, it plateaus after about 10 generations, reaching an upper bound of 60 (i.e. 60\% survival rate). 

Figure \ref{fig:coefficients_nearest} shows how the prey coefficients evolved over generations. We observe moderate to strong convergence \emph{within} ES trials for most coefficients -- except wiggle, as indicated by the exceptionally wide error bars. Fitness again appears to be the driving force behind strong ES convergence -- as converging ES oftentimes also include the fittest individual in their error margin (which is not the case for wiggle). \emph{Between} ES trials, on the other hand, we only see consistent convergence across ES trials for alignment, separation, dodge, and repel. Especially the agreement for repel is striking, as ES strongly converge both \emph{within} as well as \emph{between} trials. Notably, repel converges to a particularly low coefficient value of $[0.0,0.2]$ -- suggesting that this is a crucial requirement for prey survival under the attack-nearest strategy. Cohesion and wiggle exhibited more variability and sparser distributions at the final generation -- indicating a less pronounced or more context-dependent influence on prey survival. When visually inspecting the fittest simulation of each ES trial, we observe that for three out of five ES trials, the evolved behaviour was to stay in perfect starting formation -- cruising straight ahead, without reacting to the predator, resulting in equidistant prey which were being eaten one-by-one until the end of the simulation. 

\subsection{Attack-Random Hunting Strategy ($\mathbf{\hat{a}}(B)=\mathbf{\hat{a}_{attr}}(B)$)}

\begin{figure}[h!]
    \centering
    \includegraphics[width=\linewidth]{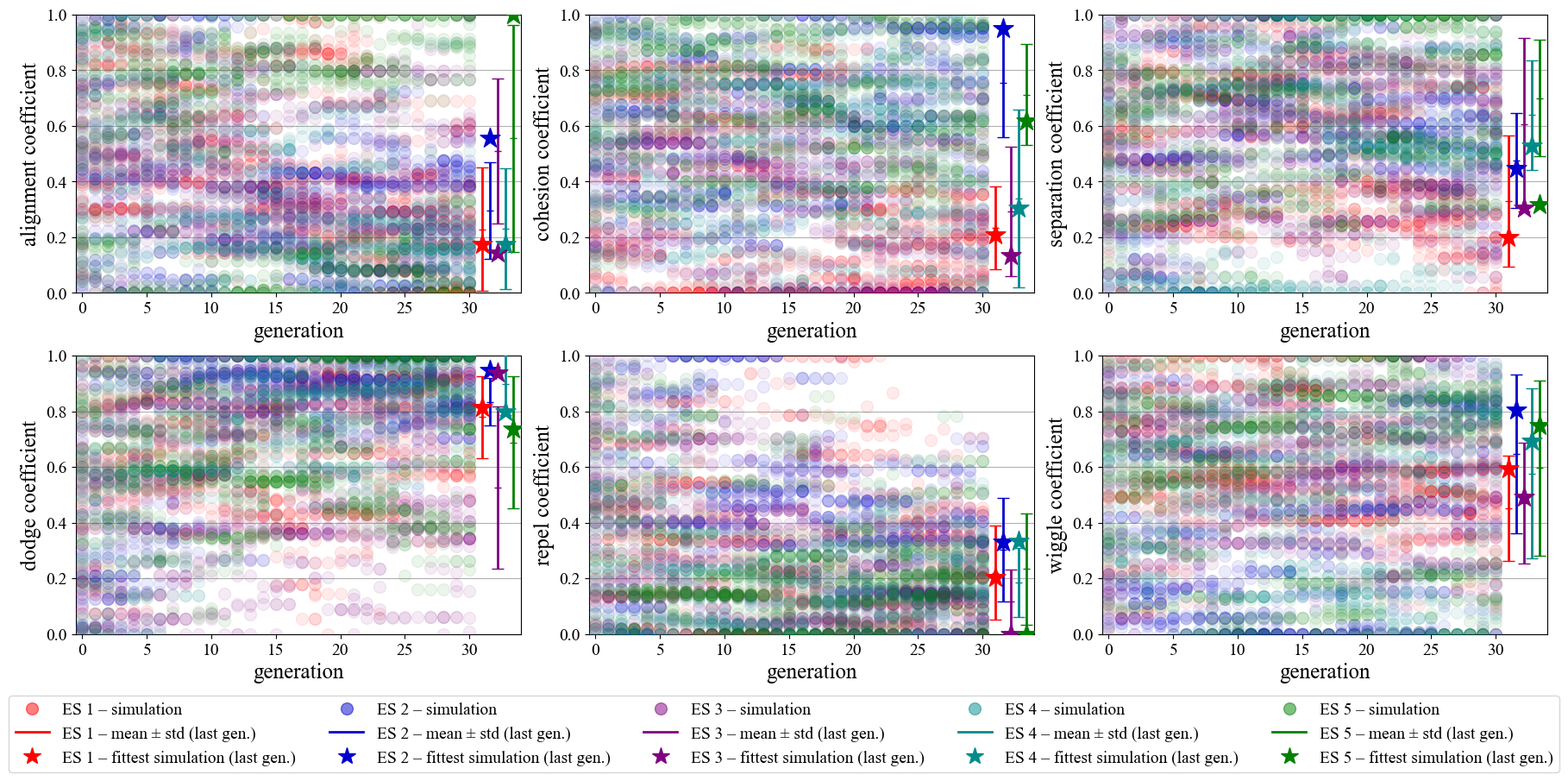}
    \caption{From left to right, top to bottom: evolution of prey coefficients $gene(s):=\langle c_{ali}, c_{coh}, c_{sep}, c_{dod}, c_{rep}, c_{wig}\rangle$ under the attack-random hunting strategy, $\mathbf{\hat{a}}(B)=\mathbf{\hat{a}_{attr}}(B)$, for five different ES trials -- corresponding to five different colours. Each dot (\CIRCLE) corresponds to an individual simulation, where overlapping simulations result in darker-shaded dots. The vertical bars indicate the population's error margin (mean $\pm$ std) at the last generation of each ES, whereas the stars ($\bigstar$) indicate the coefficient's value of the fittest individual of the last generation.}
    \label{fig:coefficients_random}
\end{figure}

\noindent For the attack-random hunting strategy $\mathbf{\hat{a}}(B)=\mathbf{\hat{a}_{attr}}(B)$, the fitness improvement is less pronounced than for the aforementioned attack-nearest strategy (see Figure \ref{fig:fitness_all}) -- showing a near-negligible fitness improvement. Figure \ref{fig:coefficients_random} demonstrates how the prey coefficients evolved over generations. We observe moderate convergence \emph{within} ES trials for all coefficients -- as confirmed by their medium-sized error margins. We furthermore see that \emph{between} ES, the coefficients only converge towards similar value ranges for the escape tendencies -- i.e., dodge, repel and wiggle. It is worth noting that repel again shows preference for lower values in $[0.0,0.4]$, whereas the other two tend to be in the medium-to-high range (dodge in $[0.6,1.0]$ and $[0.4,0.9]$). For the flocking tendencies, i.e., alignment, cohesion and separation, there is higher variability and no clear agreement across ES trials. Visual inspection of the fittest simulation of each ES trial again shows a similar behaviour as for attack-nearest -- with prey mostly staying in their initial formation. However, behaviour is more variable across prey, and a higher tendency for predator avoidance can be observed. 

\subsection{Attack-Peripheral Hunting Strategy ($\mathbf{\hat{a}}(B)=\mathbf{\hat{a}_{attp}}(B)$)}

\begin{figure}[h!]
    \centering
    \includegraphics[width=\linewidth]{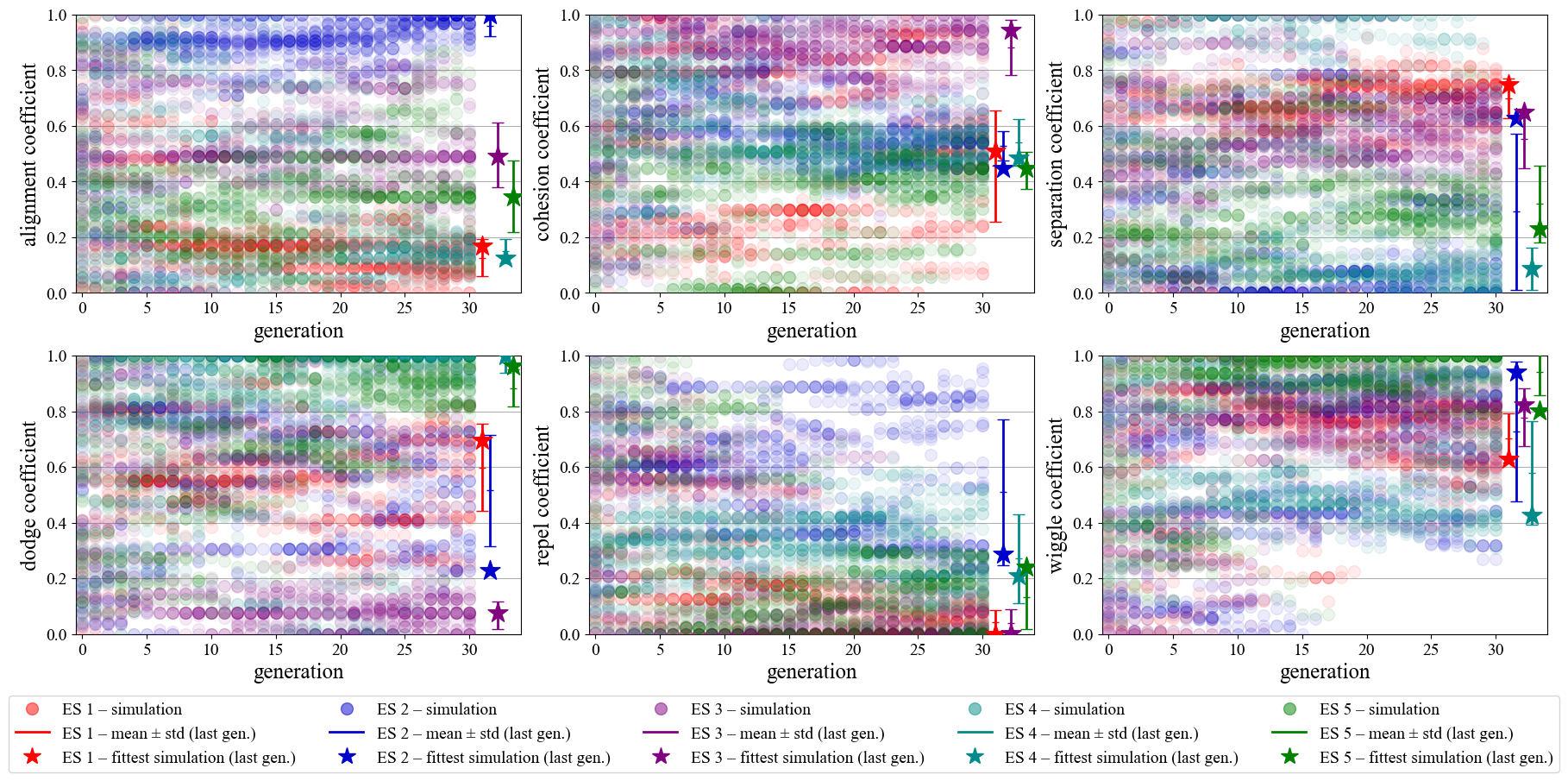}
    \caption{From left to right, top to bottom: evolution of prey coefficients $gene(s):=\langle c_{ali}, c_{coh}, c_{sep}, c_{dod}, c_{rep}, c_{wig}\rangle$ under the attack-peripheral hunting strategy, $\mathbf{\hat{a}}(B)=\mathbf{\hat{a}_{attp}}(B)$, for five different ES trials -- corresponding to five different colours. Each dot (\CIRCLE) corresponds to an individual simulation, where overlapping simulations result in darker-shaded dots. The vertical bars indicate the population's error margin (mean $\pm$ std) at the last generation of each ES, whereas the stars ($\bigstar$) indicate the coefficient's value of the fittest individual of the last generation.}
    \label{fig:coefficients_peripheral}
\end{figure}

\noindent The fitness distribution for the attack-peripheral hunting strategy ($\mathbf{\hat{a}}(B)=\mathbf{\hat{a}_{attp}}(B)$) shows a comparative spread, where the fitness distribution ranges from 52 to 91 (i.e. 9 to 48 preys killed in the simulation's runtime). Despite this variability, a slight fitness improvement is achieved as the ES trials evolve. Figure \ref{fig:coefficients_peripheral} shows how the coefficients evolved over generations. We again observe moderate to strong \emph{within}-ES convergences for all coefficients -- where alignment and cohesion show the most narrow error margins. Regarding the \emph{between}-ES agreement of the converged-to values, however, we observe a large variability across ES for most coefficients -- as only repel shows particular preference for low ($[0.0,0.4]$) values, and wiggle for higher ($[0.5,1.0]$) values. The remaining coefficients all vary considerably in their convergences across ES runs. When visually assessing the ES trial's fittest simulations, we see a very similar behaviour to the attack-random results -- with prey generally keeping their formation, but being less constrained than for the attack-nearest simulations. Further, we can clearly observe wiggling behaviour, which was not the case for any of the other attack strategy results.

\section{Discussion}
\label{Discussion}
We investigated which relative contributions of six prey behaviours -- alignment ($c_{ali}$), cohesion ($c_{coh}$), separation ($c_{sep}$), dodge ($c_{dod}$), repel ($c_{rep}$), and wiggle ($c_{wig}$) --  prey boids converge to, for a given predator strategy, using an ES. We found that the ES was successful in maximising fitness, and converging on specific coefficients, typically driven by local improvements in the fitness landscape.  
The behaviours that evolved for the different predator attack strategies are largely similar, except for the attack-centroid strategy -- which favours more individualistic predator-escaping behaviour -- as compared to the other three strategies -- which favour flocking tendencies more. Still, as Section \ref{results} demonstrated, each strategy favoured a different combination of coefficients and differed in terms of convergence \emph{within} and \emph{between} ES trials. Only the ES runs for the attack-nearest strategy showed a clear improvement in fitness. For all other strategies, no relevant improvements in fitness could be observed. This can be explained by the properties of the different attack strategies. The attack-nearest strategy resulted in the lowest survival rate of the prey, which is in line with our hypothesis, and furthermore agrees with previous research that found this strategy to be the most efficient \citep{Ojo_2023, vonmoll2016evolutionary}. The optimal prey survival behaviour under this hunting strategy is to ``do nothing'' -- that is, to stay in the initial starting formation, which perfectly equispaces the prey -- preventing the predator from eating more than one prey at a time (which is a possible scenario in our simulations). Since the simulation is run for a fixed time and the predator is faster than the prey, prey essentially ``wait-out'' the end of the simulation. This generates a very reliable fitness. Reacting to the predator using the escape tendencies can result in higher fitness, but not reliably so -- as by chance in some simulations, multiple clustered prey could be eaten by the predator in one go. In a sense, this shows how environmental context influences optimal behaviour \citep{palmer2021reactive}. 

Similarly, the evolved behaviours for the other attack strategies can be understood. The attack-centroid strategy resulted in the highest survival rate for the prey -- again aligning with our hypothesis and previous work \citep{Demsar2014-km}. This can be easily explained by considering that the flock's centroid typically does not coincide with an individual prey -- resulting in the predator aiming to catch `empty space'. As a consequence, there is less pressure for the prey to flock, as long as they avoid the predator -- as indicated by the relatively high escape coefficients. Since many possible combinations of escape behaviours can be successful for this strategy, the random initialization and all consecutive generations achieve a very high average fitness, that can hardly be improved upon due to the stochasticity inherent in both the boid's model and ES. 

The attack-random and attack-peripheral strategies place somewhere in between attack-nearest and attack-centroid -- both in terms of highest achievable fitness for the prey, as well as the evolved behaviours. Both strategies are more efficient than attack-centroid, since they target and hunt actual, individual prey, but fall short of attack-nearest -- which targets the \emph{closest} prey, as observed by \citet{Ojo_2023}. This is reflected in the ES achieving maximal fitness values between the two extreme strategies. Both attack-random and attack-peripheral are rather unpredictable. For the prey, the evolved behaviour consists of generally staying in flock formation, which aligns with the findings of \citep{Kunz2006-ze}. They turn away from the predator if it is spotted, however, they avoid sharp escape manoeuvres that break flock formation. This exploits the inefficient hunting of the predator -- which regularly chooses prey that are further away -- leading to long travel times during which prey are only caught by chance. 

Altogether, in terms of evolved coefficients over all hunting strategies, we can highlight a general tendency towards a relatively low repel coefficient across the ES trials -- while other coefficients did not present the same consistency. Other, more subtle patterns are a relatively high dodge and wiggle coefficient -- which are deemed beneficial to avoid predators' attacks. This contrasts with the findings from \citet{Ojo_2023}, who found position-based escape (repel) to perform better than direction-based escape (dodge). This deviation could potentially be attributed to the interaction between coefficients in our research, while \citet{Ojo_2023} only investigated behaviours in isolation. No clear pattern can be identified for the boids coefficients, as multiple coefficient combinations result in the observed flocking behaviour.

\subsection{Limitations \& Future Work}
Our study has several limitations, which also pave the way for promising directions for future work. We highlight the following:
\begin{itemize}
    \item \emph{Environment Setting}: The simulation run in our experiment assumed a 2D toroidal environment with limited size. In future research, the environment could be expanded, in terms of both size and dimensions, potentially removing the toroidal boundaries to better approximate open or higher-dimensional spaces. Furthermore, the addition of obstacles or other environmental features may vary the behaviour of the flock, thus possibly leading to interesting findings.
    \item \emph{Initialisation and stochasticity}: In order to account for stochasticity and isolate the effects of the behavioural coefficients, every simulation has been run starting from the exact same initial position and direction of the boids. However, to enhance the realism of those simulations and introduce stochasticity, different initial setups could be explored.
    \item \emph{Computational Cost}: Despite the extensive use of parallelisation in the experiment implementation, the demanding computational cost of the simulations prohibited long ES runs, and constrained our research to test ES for only $\mathcal{N}_g=30$ generations (whereas most of the other studies run their GA for 100 or more \citep{olsen2018genetic, Kunz2006-ze}). With more computational resources, a more extensive study would be recommended to effectively assess the convergence of the ES.
    \item \emph{Parameters}: Given the broad parameter space, there is significant potential for further exploration. While this study fixed most parameters to maintain a realistic simulation consistent with the chosen hunting strategy, future research could examine a wider array of strategies and parameter settings, and/or combine hunting strategies, such as first attacking the centre of a flock and following up with attacking nearest prey \citep{Demsar2014-km}. Also, introducing predator collaboration in the hunting tactic by means of increasing the number of predators, could lead to interesting prey dynamics.
\end{itemize}

\section{Conclusion}
We investigated to what movement behaviour the prey evolve to, in terms of relative contributions of alignment, cohesion, separation, dodge, repel and wiggle tendencies, to maximise their collective survival -- under different predator hunting strategies. From the prey perspective, we conclude that flocking (relating to the coefficients of alignment, cohesion and separation) is usually beneficial, whereas the optimality of escaping (relating to the coefficients of dodge, repel and wiggle) slightly varies according to the hunting strategy adopted by the predator. From the perspective of the predators, we can conclude that attacking the centroid of the prey flock, or attacking a randomly-selected prey, is not an effective hunting strategy, while attacking the nearest prey is reliably most successful. This research serves to improve our understanding of the complex interactions between prey and predator behaviours. We encourage future research to gain deeper insights into the dynamics that were presented here.


\bibliographystyle{apalike}
\bibliography{references}

\newpage
\appendix
\section{Fixed Parameter Settings}
\label{app:fixed_parameter_settings}

We selected the parameter settings in Table \ref{tab:fixed_parameter_settings} to be fixed during our main experiments. 

\begin{table}[h]
\centering
\begin{tabular}{|l|l|cc|}
\hline
\textbf{Parameter} & \textbf{Explanation} & \multicolumn{2}{c|}{\textbf{Value}} \\
\hline
\hline
\multicolumn{4}{|l|}{\textbf{Boids Model}} \\
\hline
$S_x\times S_y$   & field size (px)   & \multicolumn{2}{c|}{$1920 \times 1080$} \\
\cline{3-4}
 & & \textbf{Predator} & \textbf{Prey}\\
 \cline{3-4}
 $M,N$         & number of boids            & 1  & 100 \\
$r_P^{pred,prey}$      & perception radius (px)       & 3000 &\textcolor{lightgray}{\{300, }750\textcolor{lightgray}{, 1500, 3000\}} \\
$r_S^{pred,prey}$      & separation radius (px)       & 100  & 50 \\
$fov^{pred,prey}$      & field of view, per eye ($\degree$)      & $\frac{120}{2}$  & \textcolor{lightgray}{\{$\frac{60}{2}$, $\frac{120}{2}$, }$\frac{240}{2}$\textcolor{lightgray}{, $\frac{360}{2}$\}} \\
$\theta^{pred,prey}_{max}$      & maximum rotation angle ($\degree$)      & 90  & 90 \\
$a^{pred,prey}$      & acceleration (px/step$^2$)      & 5000  & 2500 \\
$v^{pred,prey}$        & velocity (px/step)       & 500  & 200  \\
$\theta_{wig}$      & wiggle angle ($\degree$)      &    & 30 \\
$f_{wig}$      & wiggle frequency (rad/step)      &    & 14 \\
\hline
\hline
\multicolumn{4}{|l|}{\textbf{Evolutionary Strategy (ES)}} \\
\hline
fps      & simulation rate (fps)  & \multicolumn{2}{c|}{100} \\
$\mathcal{T}$   & duration of a simulation (steps)  & \multicolumn{2}{c|}{2000} \\
$T$   & duration of a simulation (s)  & \multicolumn{2}{c|}{20} \\
$\mathcal{N}_s$   & number of simulations per generation, population size   & \multicolumn{2}{c|}{\textcolor{lightgray}{\{}30\textcolor{lightgray}{, 50\}}} \\
$\mathcal{N}_g$   & number of generations    & \multicolumn{2}{c|}{30} \\ 
$\mu_e$   & elite size    & \multicolumn{2}{c|}{2} \\
$\mu$   & mutation rate    & \multicolumn{2}{c|}{\textcolor{lightgray}{\{0.05, }0.3\textcolor{lightgray}{, 0.7\}}} \\
\hline
\end{tabular}
\caption{Fixed parameter settings relating to the Boids Model and the evolutionary strategy (ES) that were used for running the main experiments. The settings were tweaked such, that they produced naturalistic predator and prey motion dynamics, and furthermore allowed ES convergence over generations. The parameter values in light grey correspond to the settings examined in the sensitivity analyses (see Appendix \ref{app:sensitivity_analyses}), but not selected for running the main experiments.}
\label{tab:fixed_parameter_settings}
\end{table}

\begin{wrapfigure}[15]{r}{0.45\textwidth}
    \centering
    \vspace{-11pt} 
    \includegraphics[width=0.45\textwidth]{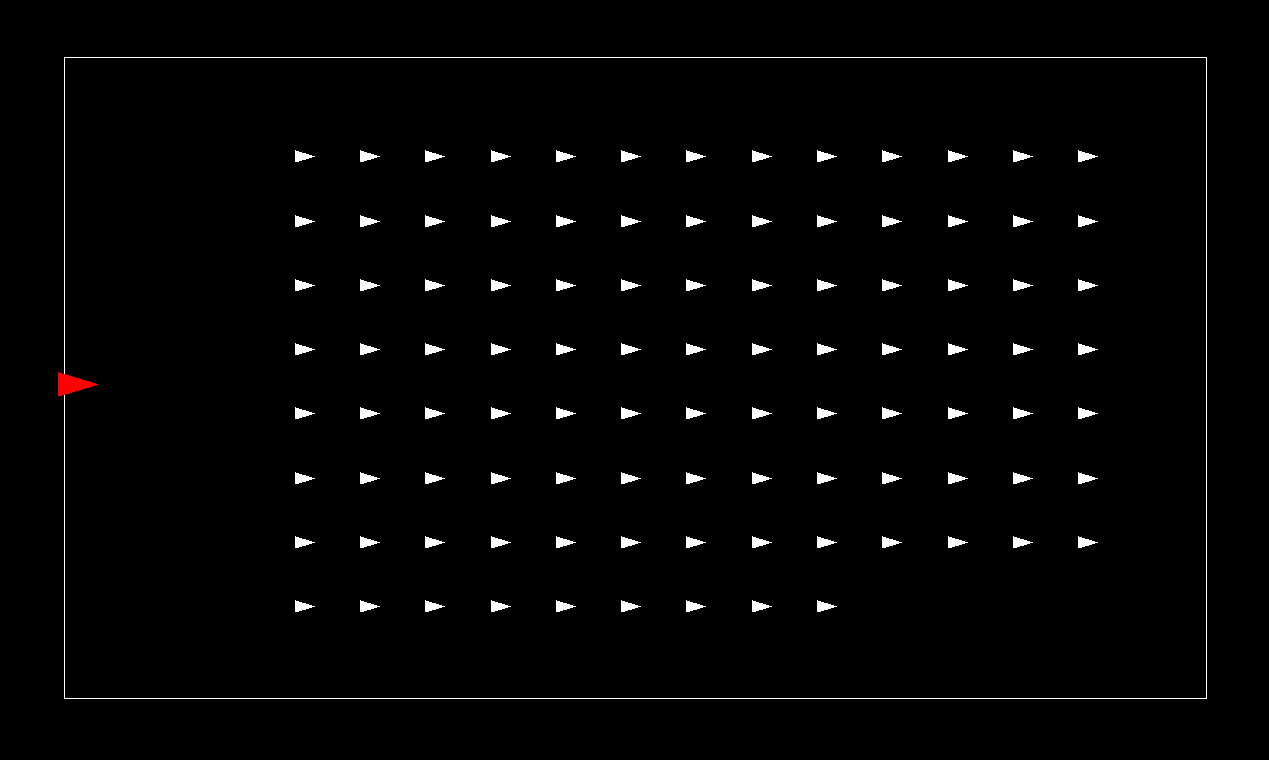}
    \caption{Initial layout of the predator (red) and prey (white) that was fixed for every simulation.}
    \label{fig:initial_layout}
    \vspace{0pt} 
\end{wrapfigure}
\noindent Relating to the Boids Model, $S_x\times S_y$ ensured sufficient space for the boids to move around, without overcrowding the field. We chose to include one single predator ($M=1$) to better isolate and examine the effects of its hunting strategy on the prey, and to avoid confounding prey behaviour with responses to multiple approaching predators. The choice for $N=100$ prey was in line with related literature \citep{Kunz2006-ze}, being an intuitive number to work with, and practical for fitness quantification. The parameters relating to the boids' motion dynamics, $r_P^{pred,prey}$, $r_S^{pred,prey}$, $fov^{pred,prey}$, $\theta_{max}^{pred,prey}$, $a^{pred,prey},v^{pred,prey}$ and wiggle parameters $\theta_{wig},f_{wig}$ were tuned such, to produce naturalistic predator and prey motion dynamics. The specific settings of $r_P^{prey}$ and $fov^{prey}$ were further motivated by promoting the stability of ES convergence (see Appendix \ref{app:sensitivity_analyses}).

Furthermore, for every simulation, we fixed the initial layout of the predator (in red) and prey (in white) to be as displayed in Figure \ref{fig:initial_layout}. Taking the predator's field of view ($fov^{pred}$) and the toroidal boundary conditions of the field into account, we opted for this layout to ensure that the predator has sufficient candidate prey within sight, such that its hunting strategy directly reflects in its movement behaviour. Furthermore, the evenly-spaced prey grid allows sufficient freedom for flocking behaviour to emerge. Taken together, this layout ensures that most is made out of every simulation, within duration $\mathcal{T}$. It is worth noting that, although the initial conditions of each simulation are deterministic, their unfolding is inherently stochastic.

For ES, we fixed $\text{fps}$ and $\mathcal{T}$ (in steps\footnote{$\mathcal{T}$ was fixed in terms of steps, not seconds, as the simulation duration in terms of steps is not subject to computational power.}) such, that any change in the simulation's prey coefficients $gene(s)$ were allowed enough time to have a noticeable impact on the simulation's fitness. Within $\mathcal{T}$, this required a considerable amount of prey to be killed when displaying unsuccessful flocking behaviour, and conversely, sufficient chances to escape from the predator when employing successful flocking behaviour. Bearing computational time in mind, we set $\mathcal{N}_s$ and $\mathcal{N}_g$ such, that the ES showed to converge (see Appendix \ref{app:sensitivity_analyses}). Relating to the generational updating, $\mu_e$ and $\mu$ were set such, to result in a balance between exploration and exploitation (see Appendix \ref{app:sensitivity_analyses}).

\section{Sensitivity Analyses}
\label{app:sensitivity_analyses}
To scrutinise the impact of varying the prey perception radius $r_P^{prey}$, field of view $fov^{prey}$, number of simulations per generation $\mathcal{N}_s$ and mutation rate $\mu$ on the stability of ES convergence, we performed sensitivity analyses on the aforementioned parameters. The experiments were conducted under the predator's attack-nearest hunting strategy ($\mathbf{\hat{a}}(B)=\mathbf{\hat{a}_{attn}}(B)$), for $\mathcal{N}_g=20$ generations, keeping all other parameters fixed to the settings in Table \ref{tab:fixed_parameter_settings}. For each parameter, the setting that yielded the best results was selected to be held fixed in the main experiments. 

Favourable parameter settings are characterised by the population's ability to `learn' how to survive. In other words, we sought parameter settings that resulted in a rise of the simulations' mean fitness over generations -- corresponding to stable convergence of the simulations' prey coefficients towards the fittest simulation's value. 

\subsection{Perception Radius ($r_P^{prey}$)}
\begin{figure}[htbp]
    \centering
    \begin{subfigure}[t]{\textwidth}
        \centering
        \begin{minipage}[t]{0.24\textwidth}
            \centering
            \includegraphics[width=\textwidth]{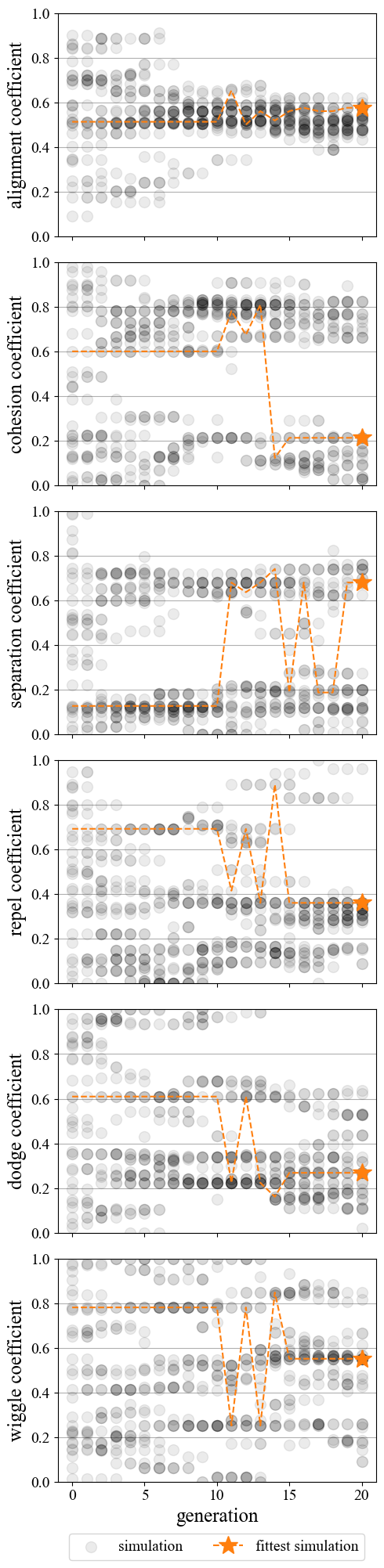}
        \end{minipage}
        \hfill
        \begin{minipage}[t]{0.24\textwidth}
            \centering
            \includegraphics[width=\textwidth]{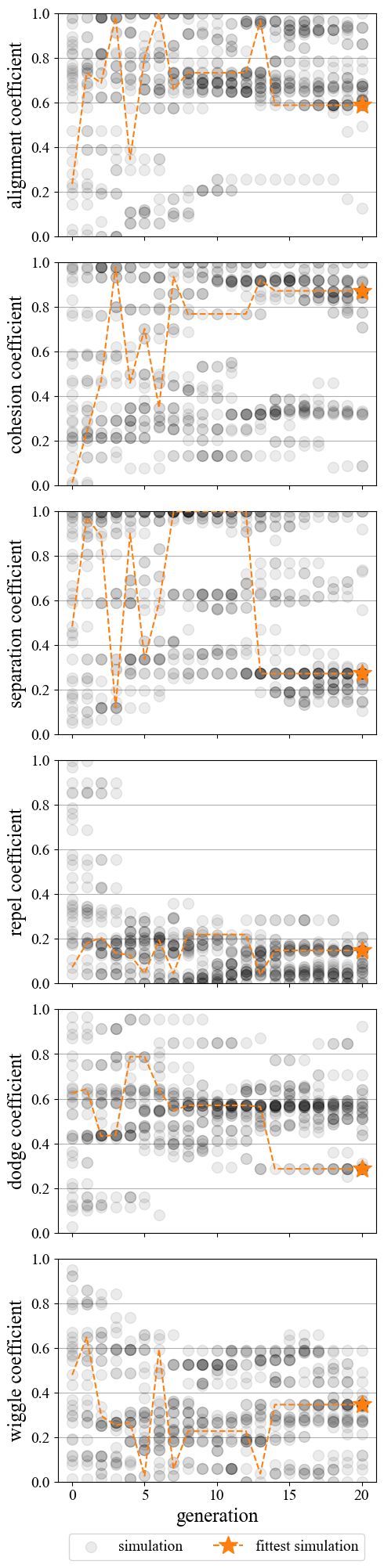}
        \end{minipage}
        \hfill
        \begin{minipage}[t]{0.24\textwidth}
            \centering
            \includegraphics[width=\textwidth]{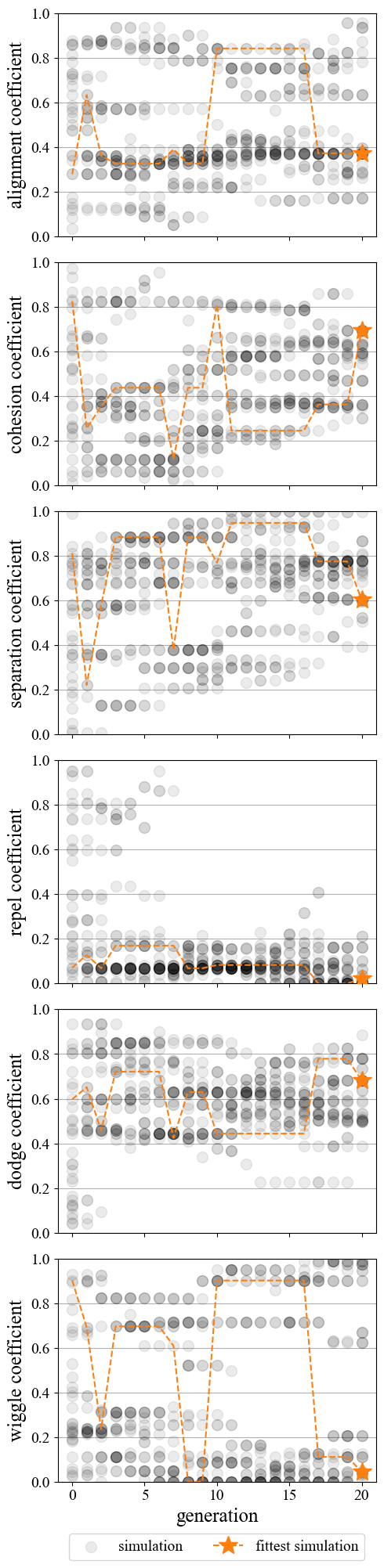}
        \end{minipage}
        \hfill
        \begin{minipage}[t]{0.24\textwidth}
            \centering
            \includegraphics[width=\textwidth]{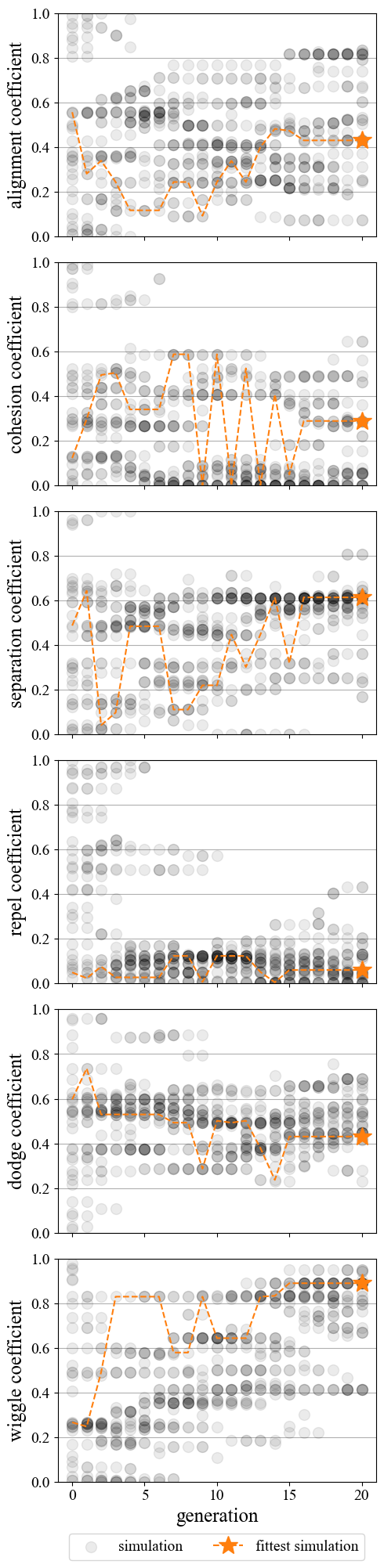}
        \end{minipage}
        \caption{Evolution of the simulations' prey coefficients across generations, under varying $r_P^{prey}$. The fittest simulation is marked as a dashed line.}
    \end{subfigure}

    \vspace{1em} 

    \begin{subfigure}[t]{\textwidth}
        \centering
        \begin{minipage}[t]{0.24\textwidth}
            \centering
            \includegraphics[width=\textwidth]{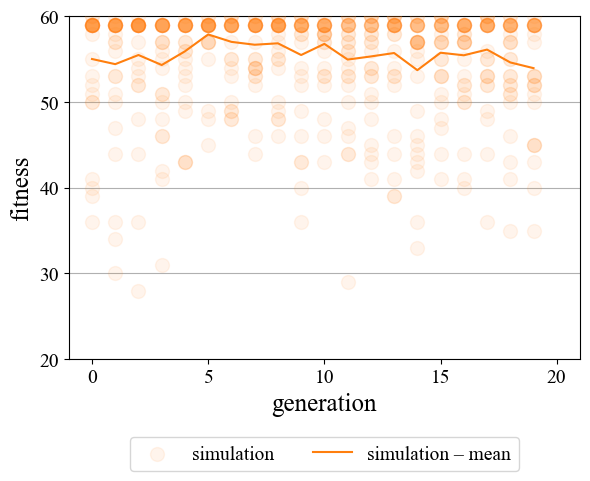}
            \smallskip
            \footnotesize $r_P^{prey} = 300$
        \end{minipage}
        \hfill
        \begin{minipage}[t]{0.24\textwidth}
            \centering
            \includegraphics[width=\textwidth]{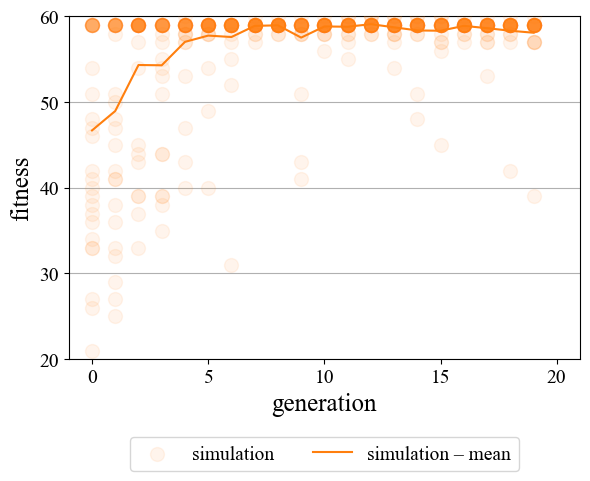}
            \smallskip
            \footnotesize $r_P^{prey} = 750$
        \end{minipage}
        \hfill
        \begin{minipage}[t]{0.24\textwidth}
            \centering
            \includegraphics[width=\textwidth]{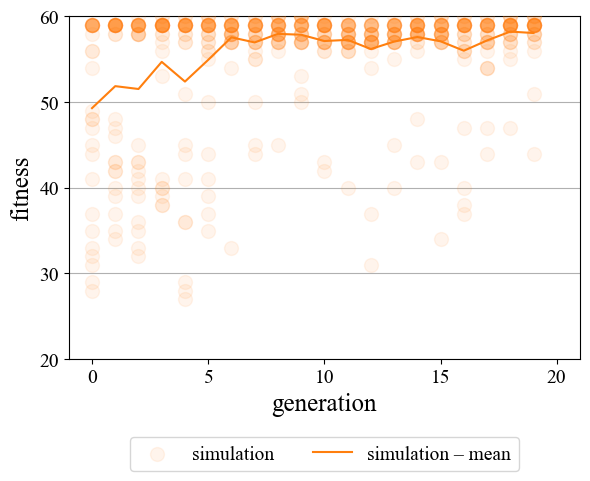}
            \smallskip
            \footnotesize $r_P^{prey} = 1500$
        \end{minipage}
        \hfill
        \begin{minipage}[t]{0.24\textwidth}
            \centering
            \includegraphics[width=\textwidth]{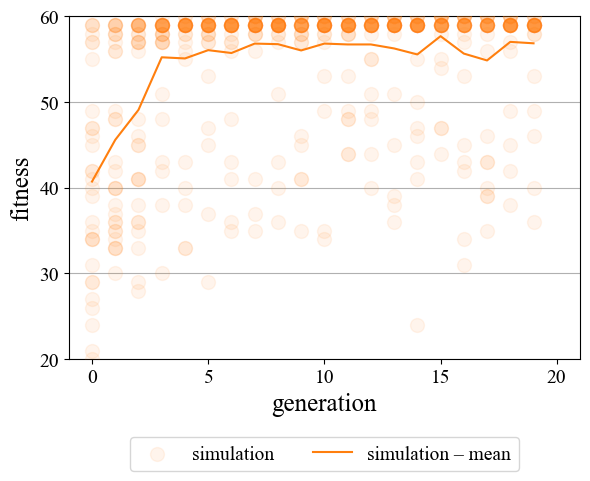}
            \smallskip
            \footnotesize $r_P^{prey} = 3000$
        \end{minipage}
        \caption{Evolution of the simulations' fitness across generations, under varying $r_P^{prey}$. The mean fitness is marked as a solid line.}
    \end{subfigure}

    \caption{Comparison of ES convergence in terms of the simulations' prey coefficients and fitnesses, when varying the prey's perception radius $r_P^{prey} \in \{300, 750, 1500, 3000\}$. The sensitivity analyses were run for $\mathcal{N}_g=20$ generations. Other parameter settings were fixed to the settings in Table \ref{tab:fixed_parameter_settings}.}
    \label{fig:sensitivity_rP}
\end{figure}

As Figure \ref{fig:sensitivity_rP} illustrates, varying the prey's perception radius across $r_P^{prey}\in\{300,750,1500,3000\}$ only partly affected the stability of ES convergence within $\mathcal{N}_g=20$. We noted that for $r_P^{prey}=300$, the distributions of prey coefficients remained rather broadly distributed (or bifurcated) as generations evolved -- reflecting a prey coefficient landscape having multiple local optima. This reflected itself as a jumpy behaviour of the fittest simulation's prey coefficients, and furthermore the inability of the population to steadily converge towards optimal prey coefficients. Together, this seemed to impede the collective fitness to `learn' to climb towards higher fitnesses. 

For $r_P^{prey}\in\{750,1500,3000\}$, on the other hand, we see a slight learning effect, as the simulations' mean fitness rises as generations unfold. Furthermore, prey coefficient convergence is seemingly more stable. Based on these results, there was no clear preference for either one of $r_P^{prey}\in\{750,1500,3000\}$, therefore we selected $r_P^{prey}=750$ for the main experiments -- having the least computational load, given that higher $r_P^{prey}$ imply more neighbours per prey.

\subsection{Field of View ($fov^{prey}$)}
\begin{figure}[htbp]
    \centering
    \begin{subfigure}[t]{\textwidth}
        \centering
        \begin{minipage}[t]{0.24\textwidth}
            \centering
            \includegraphics[width=\textwidth]{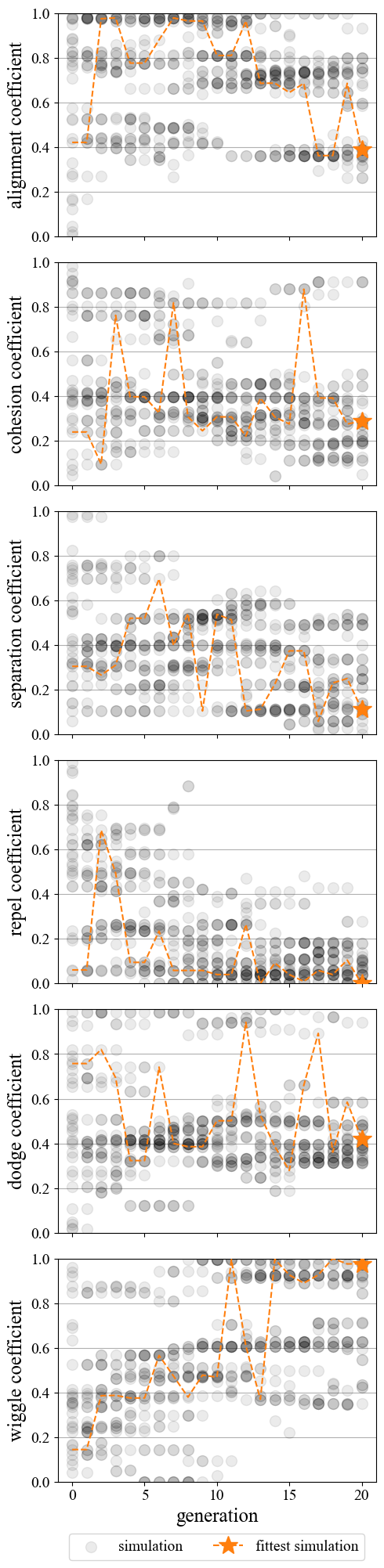}
        \end{minipage}
        \hfill
        \begin{minipage}[t]{0.24\textwidth}
            \centering
            \includegraphics[width=\textwidth]{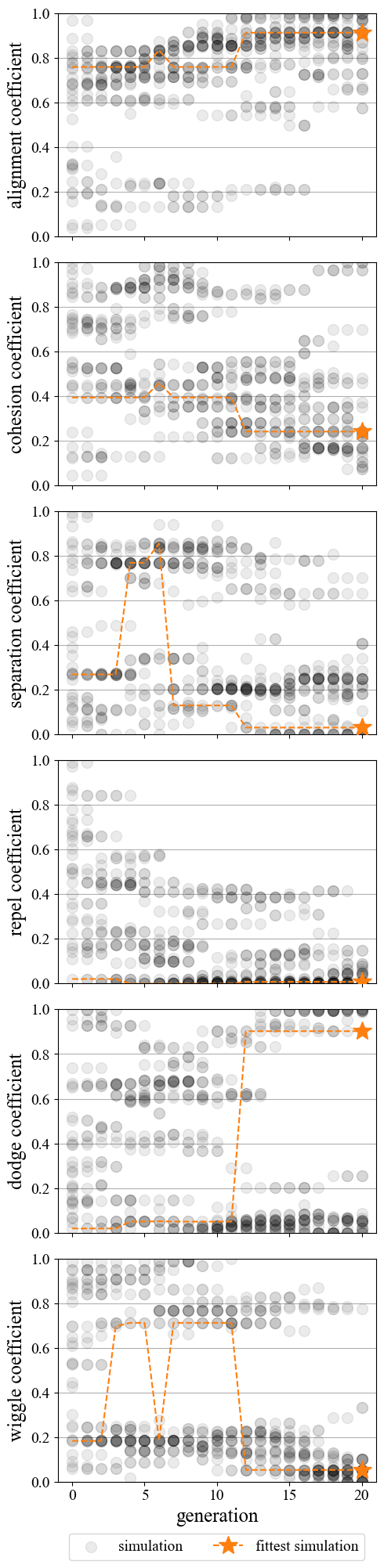}
        \end{minipage}
        \hfill
        \begin{minipage}[t]{0.24\textwidth}
            \centering
            \includegraphics[width=\textwidth]{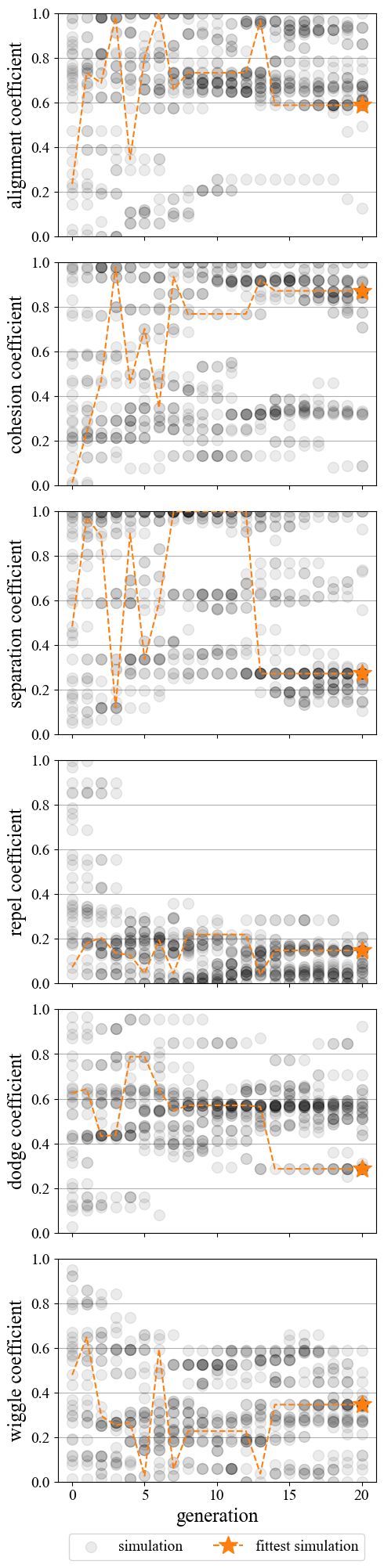}
        \end{minipage}
        \hfill
        \begin{minipage}[t]{0.24\textwidth}
            \centering
            \includegraphics[width=\textwidth]{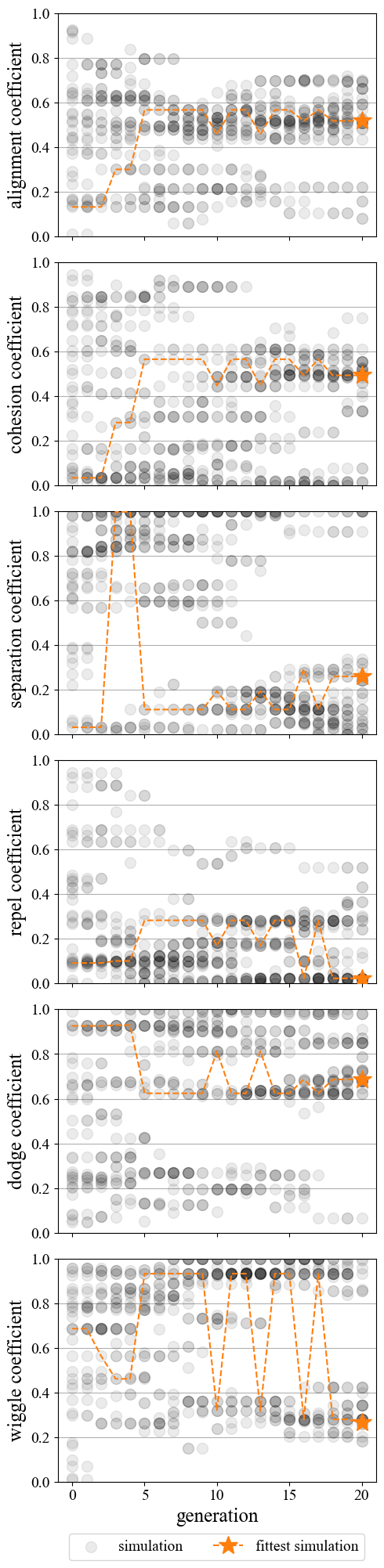}
        \end{minipage}
        \caption{Evolution of the simulations' prey coefficients across generations, under varying $fov^{prey}$. The fittest simulation is marked as a dashed line.}
    \end{subfigure}

    \vspace{1em} 

    \begin{subfigure}[t]{\textwidth}
        \centering
        \begin{minipage}[t]{0.24\textwidth}
            \centering
            \includegraphics[width=\textwidth]{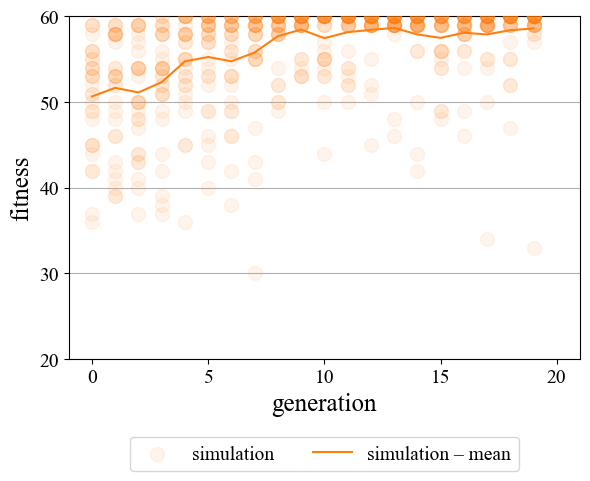}
            \smallskip
            \footnotesize $fov^{prey} = \frac{60}{2}$
        \end{minipage}
        \hfill
        \begin{minipage}[t]{0.24\textwidth}
            \centering
            \includegraphics[width=\textwidth]{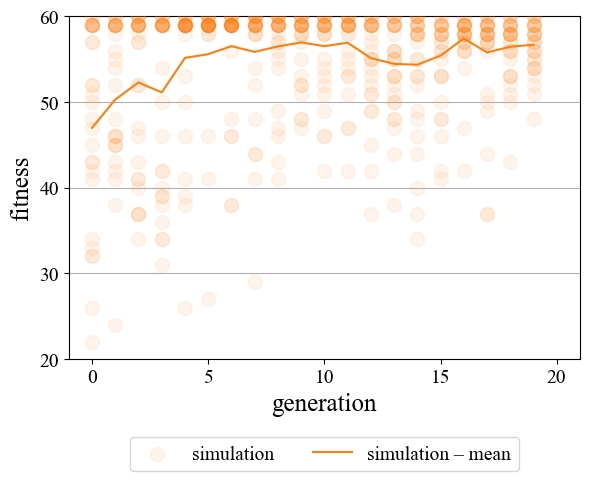}
            \smallskip
            \footnotesize $fov^{prey} = \frac{120}{2}$
        \end{minipage}
        \hfill
        \begin{minipage}[t]{0.24\textwidth}
            \centering
            \includegraphics[width=\textwidth]{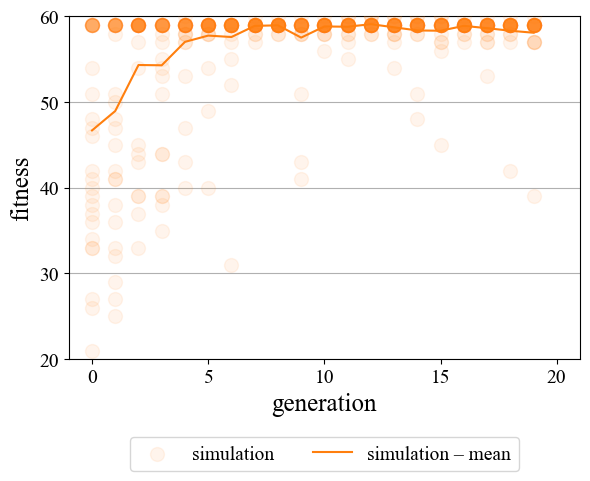}
            \smallskip
            \footnotesize $fov^{prey} = \frac{240}{2}$
        \end{minipage}
        \hfill
        \begin{minipage}[t]{0.24\textwidth}
            \centering
            \includegraphics[width=\textwidth]{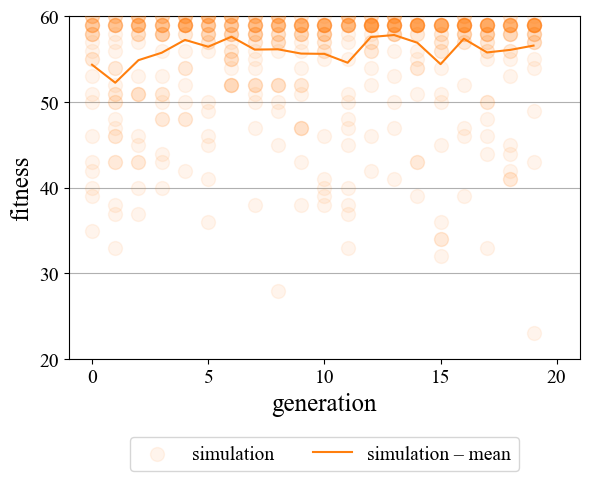}
            \smallskip
            \footnotesize $fov^{prey} = \frac{360}{2}$
        \end{minipage}
        \caption{Evolution of the simulations' fitness across generations, under varying $fov^{prey}$. The mean fitness is marked as a solid line.}
    \end{subfigure}

    \caption{Comparison of ES convergence in terms of the simulations' prey coefficients and fitnesses, when varying the prey's field of view (per eye) $fov^{prey} \in \{\frac{60}{2},\frac{120}{2},\frac{240}{2},\frac{360}{2}\}$. The sensitivity analyses were run for $\mathcal{N}_g=20$ generations. Other parameter settings were fixed to the settings in Table \ref{tab:fixed_parameter_settings}.}
    \label{fig:sensitivity_fov}
\end{figure}

As Figure \ref{fig:sensitivity_fov} illustrates, varying the prey's field of view across $fov^{prey}\in\{\frac{60}{2},\frac{120}{2},\frac{240}{2},\frac{360}{2}\}$ had a minor impact on the stability of the prey coefficients' convergence within $\mathcal{N}_g=20$. In the $fov^{prey}\in\{\frac{360}{2}$ case, we observed little-to-no rise in the simulations' mean fitness, although this was not directly reflected in the prey coefficients' convergence -- that were seemingly equally stable across all four $fov^{prey}$ scenarios. For the main experiments, we opted for $fov^{prey}=\frac{240}{2}$ to mimic nature -- in which prey typically have larger fields of view than the predator ($fov^{pred}=\frac{120}{2}$). Furthermore, this parameter setting resulted in a clear rise of the simulations' mean fitness across generations.

\subsection{Number of simulations per generation ($\mathcal{N}_s$)}
\begin{figure}[htbp]
    \centering
    \begin{subfigure}[t]{\textwidth}
        \centering
        \begin{minipage}[t]{0.24\textwidth}
            \centering
            \includegraphics[width=\textwidth]{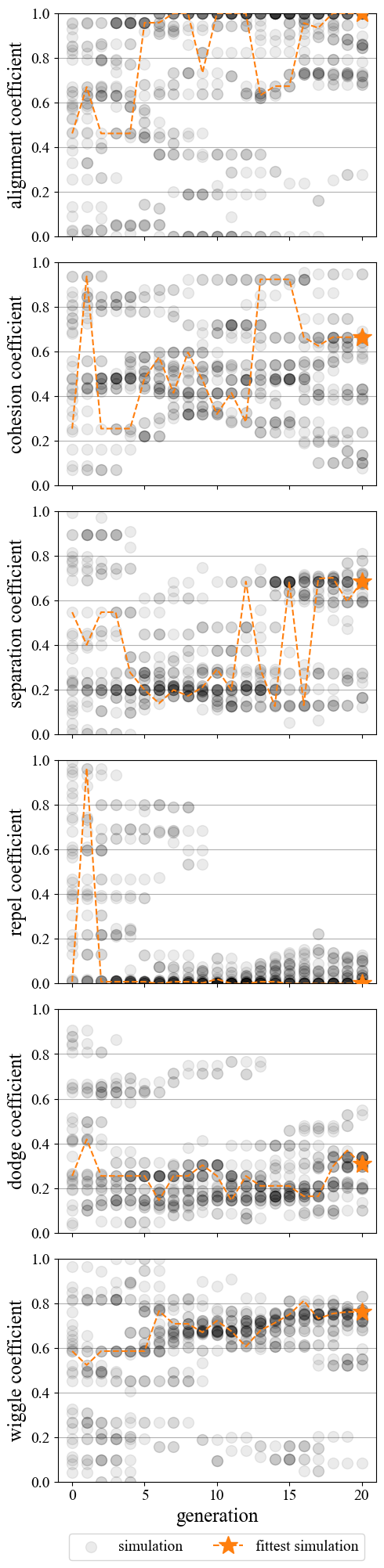}
        \end{minipage}
        \hspace{0.3cm}
        \begin{minipage}[t]{0.24\textwidth}
            \centering
            \includegraphics[width=\textwidth]{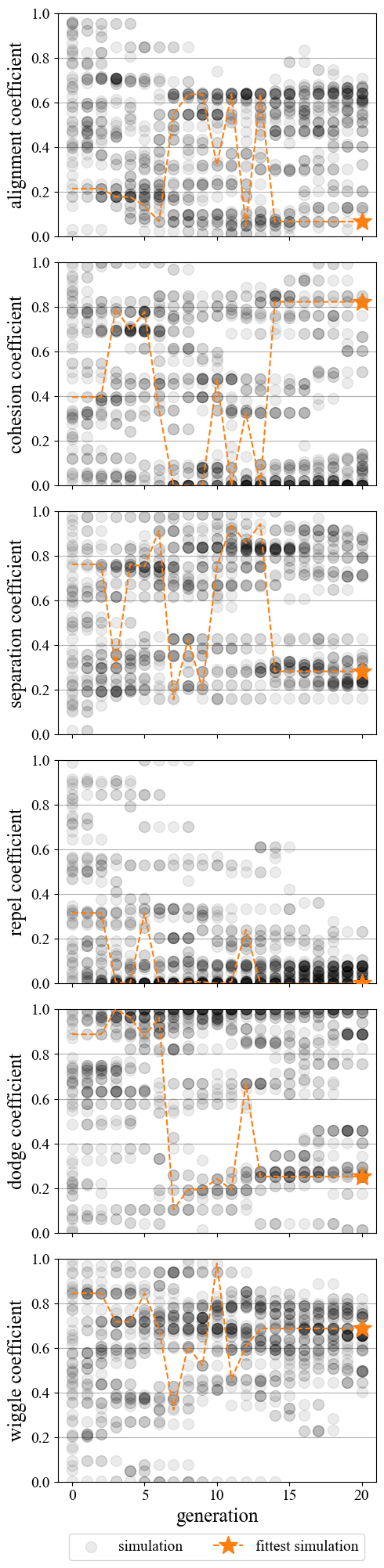}
        \end{minipage}
        \caption{Evolution of the simulations' prey coefficients across generations, under varying $\mathcal{N}_s$. The fittest simulation is marked as a dashed line.}
    \end{subfigure}

    \vspace{1em} 

    \begin{subfigure}[t]{\textwidth}
        \centering
        \begin{minipage}[t]{0.24\textwidth}
            \centering
            \includegraphics[width=\textwidth]{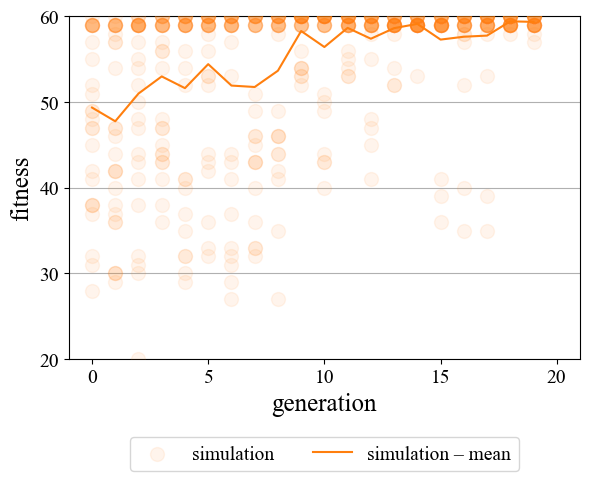}
            \smallskip
            \footnotesize $\mathcal{N}_s = 30$
        \end{minipage}
        \hspace{0.3cm}
        \begin{minipage}[t]{0.24\textwidth}
            \centering
            \includegraphics[width=\textwidth]{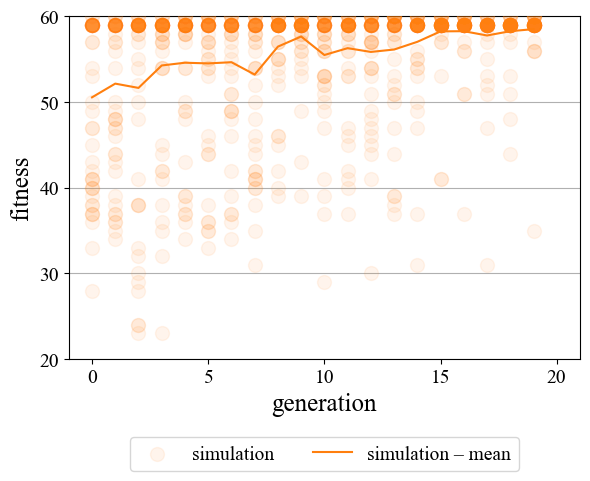}
            \smallskip
            \footnotesize $\mathcal{N}_s = 50$
        \end{minipage}
        \caption{Evolution of the simulations' fitness across generations, under varying $\mathcal{N}_s$. The mean fitness is marked as a solid line.}
    \end{subfigure}

    \caption{Comparison of ES convergence in terms of the simulations' prey coefficients and fitnesses, when varying the number of simulations per generation (population size) $\mathcal{N}_s \in \{30,50\}$. The sensitivity analyses were run for $\mathcal{N}_g=20$ generations. Other parameter settings were fixed to the settings in Table \ref{tab:fixed_parameter_settings}.}
    \label{fig:sensitivity_Ns}
\end{figure}

Looking at Figure \ref{fig:sensitivity_Ns}, varying the number of simulations per generation $\mathcal{N}_s\in\{30,50\}$ showed most favourable ES dynamics for $\mathcal{N}_s=30$. Here, the prey coefficients showed to converge more stably to the values of the fittest simulation -- whereas for $\mathcal{N}_s=50$ the distribution of the prey coefficients' values remained broad or strongly bifurcated over generations. Consequently, more simulations with suboptimal fitnesses remained within the population, whereas for $\mathcal{N}_s=30$ they had almost all reached the fitness ceiling. For these reasons, we selected $\mathcal{N}_g=30$ for our main experiments.

\subsection{Mutation rate ($\mu$)}
\begin{figure}[htbp]
    \centering
    \begin{subfigure}[t]{\textwidth}
        \centering
        \begin{minipage}[t]{0.24\textwidth}
            \centering
            \includegraphics[width=\textwidth]{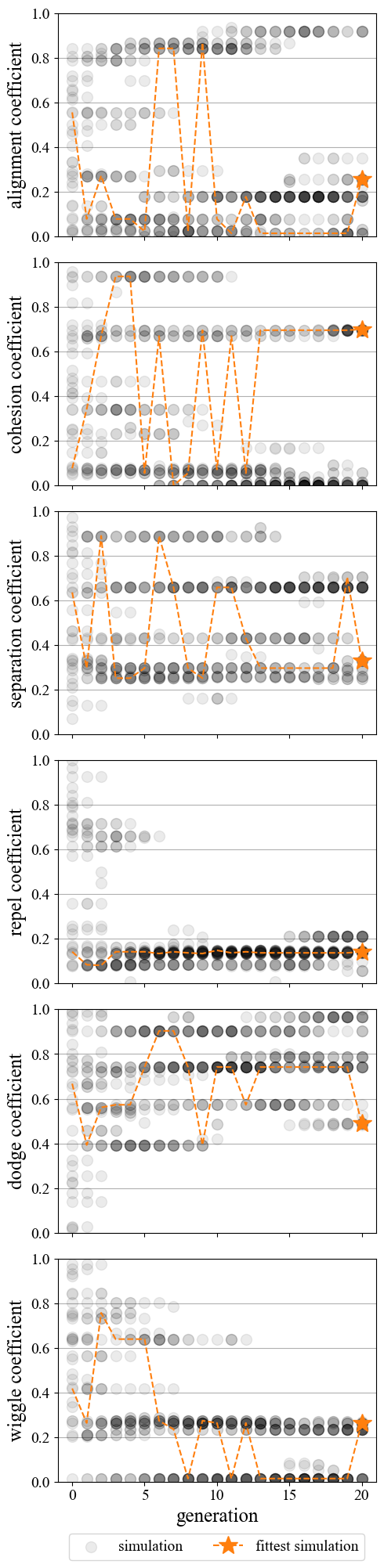}
        \end{minipage}
        \hspace{0.3cm}
        \begin{minipage}[t]{0.24\textwidth}
            \centering
            \includegraphics[width=\textwidth]{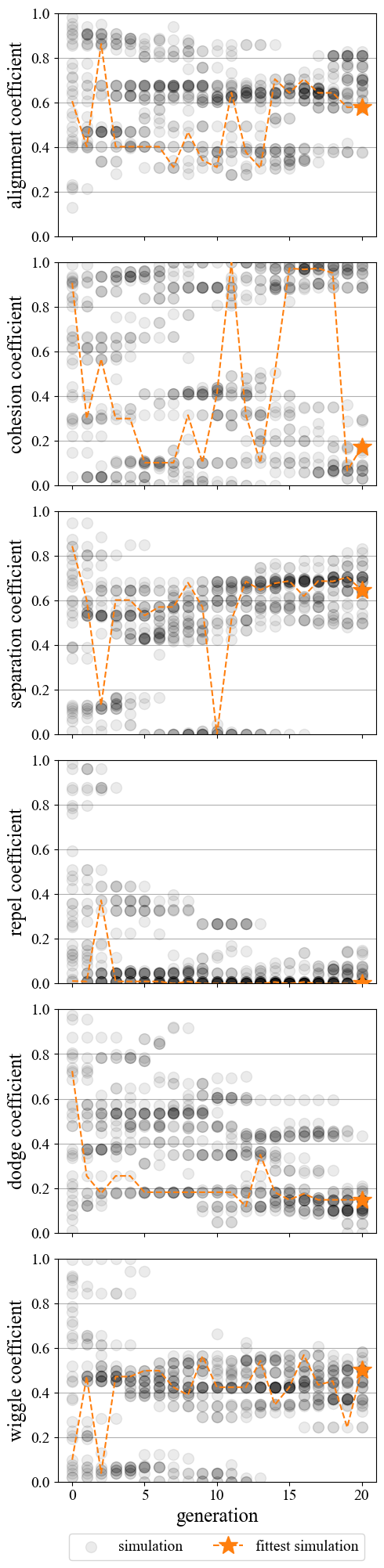}
        \end{minipage}
        \hspace{0.3cm}
        \begin{minipage}[t]{0.24\textwidth}
            \centering
            \includegraphics[width=\textwidth]{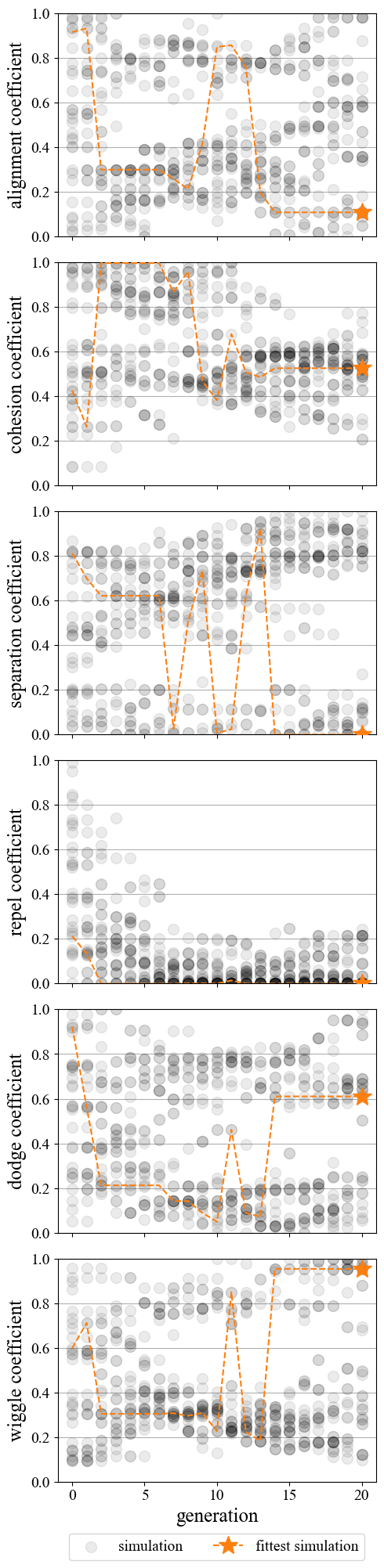}
        \end{minipage}
        \caption{Evolution of the simulations' prey coefficients across generations, under varying $\mu$. The fittest simulation is marked as a dashed line.}
    \end{subfigure}

    \vspace{1em} 

    \begin{subfigure}[t]{\textwidth}
        \centering
        \begin{minipage}[t]{0.24\textwidth}
            \centering
            \includegraphics[width=\textwidth]{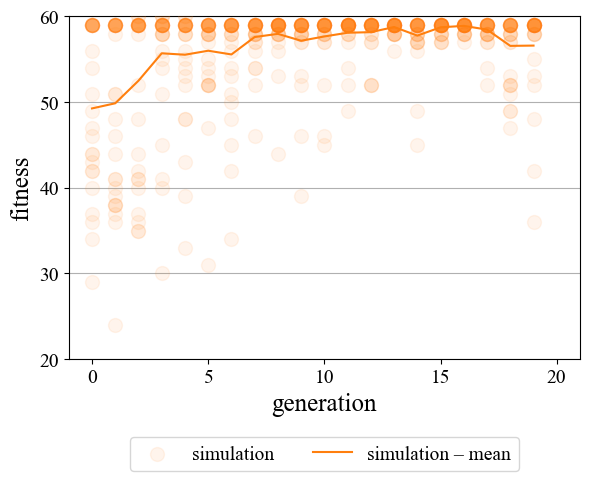}
            \smallskip
            \footnotesize $\mu=0.05$
        \end{minipage}
        \hspace{0.3cm}
        \begin{minipage}[t]{0.24\textwidth}
            \centering
            \includegraphics[width=\textwidth]{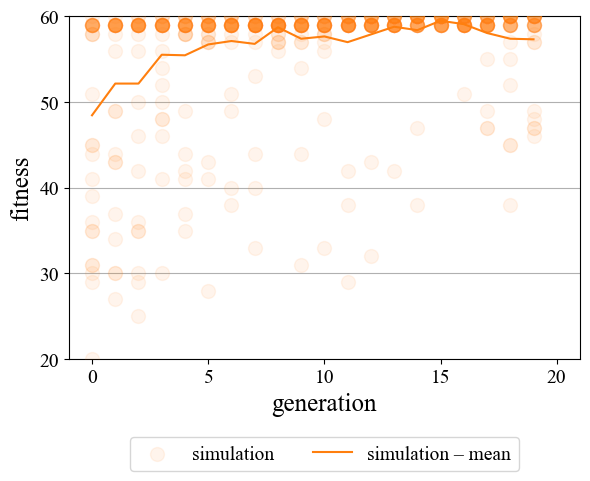}
            \smallskip
            \footnotesize $\mu=0.3$
        \end{minipage}
        \hspace{0.3cm}
        \begin{minipage}[t]{0.24\textwidth}
            \centering
            \includegraphics[width=\textwidth]{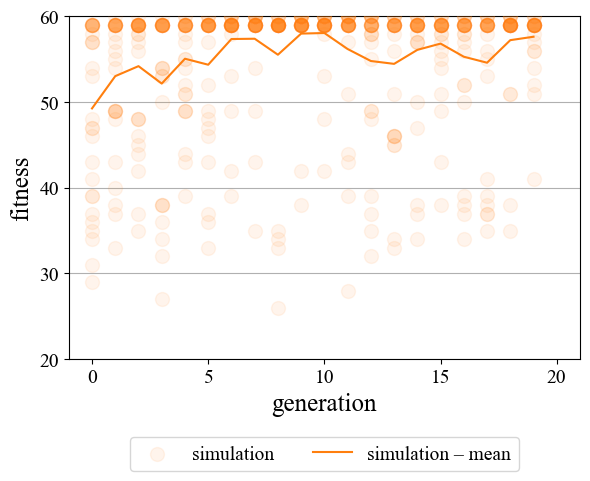}
            \smallskip
            \footnotesize $\mu=0.7$
        \end{minipage}
        \caption{Evolution of the simulations' fitness across generations, under varying $\mu$. The mean fitness is marked as a solid line.}
    \end{subfigure}

    \caption{Comparison of ES convergence in terms of the simulations' prey coefficients and fitnesses, when varying the mutation rate $\mu \in \{0.05,0.3,0.7\}$. The sensitivity analyses were run for $\mathcal{N}_g=20$ generations. Other parameter settings were fixed to the settings in Table \ref{tab:fixed_parameter_settings}.}
    \label{fig:sensitivity_mu}
\end{figure}

Looking at Figure \ref{fig:sensitivity_mu}, varying the mutation rate $\mu\in\{0.05, 0.3, 0.7\}$ showed stronger exploitation effects $\mu=0.05$ -- as in this scenario, generational variations through mutation were limited, we observe that the prey coefficients' distributions show little-to-no changes over generations -- as variations are almost entirely dependent on crossover with just a 5\% chance of mutation. This impedes exploration of the prey coefficients' landscape towards configurations that result in higher fitnesses. For $\mu=0.7$, we observed the contrary. Due to the 70\% chance of gene mutation, the prey coefficients struggle to converge, as their distributions remain broad over generations. Here in particular, the overemphasis on exploration withholds the mean fitness generation to steadily rise. Best prey coefficients' convergence and population fitnesses were achieved by balancing exploration and exploitation, as was the case for $\mu=0.3$. Therefore, this setting was selected for the main experiments.

\end{document}